\documentclass[useAMS, usenatbib, usegraphicx]{mn2e}

\newcommand{\msun}{M_\odot}

\usepackage{color}
\newcommand{\red}[1]{{{#1}}}

\begin{document}

\title[Star Formation in Molecule-Poor Galaxies]{The Star Formation Law in Molecule-Poor Galaxies}

\author[Krumholz]{
Mark R. Krumholz$^1$\thanks{mkrumhol@ucsc.edu}
\\ \\
$^1$Department of Astronomy \& Astrophysics, University of California, Santa 
Cruz, CA 95064 USA}

\maketitle

\begin{abstract}
In this paper, I investigate the processes that regulate the rate of star formation in regions of galaxies where the neutral interstellar medium is predominantly composed of non-star-forming H~\textsc{i}. In such regions, found today predominantly in low-metallicity dwarf galaxies and in the outer parts of large spirals, the star formation rate per unit area and per unit mass is much smaller than in more molecule-rich regions. While in molecule-rich regions the ultraviolet radiation field produced by efficient star formation forces the density of the cold neutral medium to a value set by two-phase equilibrium, I show that the low rates of star formation found in molecule-poor regions preclude this condition. Instead, the density of the cold neutral gas is set by the requirements of hydrostatic balance. Using this result, I extend the \citet{krumholz08c, krumholz09a, krumholz09b} model for star formation and the atomic to molecular transition to the molecule-poor regime. This ``KMT+" model matches a wide range of observations of the star formation rate and the balance between the atomic and molecular phases in dwarfs and in the outer parts of spirals, and is well-suited to implementation as a subgrid recipe for star formation in cosmological simulations and semi-analytic models. I discuss the implications of this model for star formation over cosmological times.
\end{abstract}


\section{Introduction}
\label{sec:intro}

Star formation in local galaxies appears to occur exclusively in the molecular phase of the interstellar medium \citep{wong02a, kennicutt07a, leroy08a, bigiel08a, bolatto11a}, a result that has been understood theoretically as arising from the correlation between the chemical state of interstellar gas and its temperature \citep{schaye04a, krumholz11b, glover12a}. Gas is only able to reach the low temperatures necessary for runaway gravitational collapse in regions that are well-shielded against the interstellar radiation field (ISRF), and in such regions the equilibrium chemical state of the hydrogen is H$_2$. In the inner parts of galaxies, this correlation with H$_2$ is accompanied by a strong lack of correlation between star formation and H~\textsc{i}. Instead, the H~\textsc{i} surface density distribution appears to saturate at maximum value regardless of the star formation rate \citep{leroy08a, bigiel08a}. \citet[hereafter KMT]{krumholz08c, krumholz09a, krumholz09b} and \citet{mckee10a} explained this saturation as a shielding effect: a certain column of H~\textsc{i} is required to block out the photodissociating effects of the ISRF and allow a transition to H$_2$, and observations of nearby molecular clouds are consistent with the predictions of this model \citep{lee12a}. Moreover, KMT predicted that H~\textsc{i} saturation column scale roughly inversely with metallicity, and subsequent observations of dwarf galaxies \citep{fumagalli10a, bolatto11a} and damped Lyman $\alpha$ absorbers \citep{krumholz09e, rafelski11a} have confirmed this prediction.

The situation is quite different in the outer parts of galaxies and in dwarf galaxies, where the ISM becomes completely dominated by H~\textsc{i} and H$_2$ fractions are small. Although star formation continues to correlate with H$_2$ down to the lowest H$_2$ columns that can be detected \citep{schruba11a, bolatto11a}, it \textit{also} begins to correlate with the total H~\textsc{i} column \citep{bigiel10a, bolatto11a}. The star formation timescales implied by these correlations, however, are quite different: while H$_2$ forms stars over a timescale of $\approx 2$ Gyr in both H$_2$-rich and H$_2$-poor regions, the correlation between H~\textsc{i} and star formation, when it is present at all, implies a star formation timescale of $\sim 100$ Gyr. Taken at face value, these two observations together would seem to imply that, in H~\textsc{i}-dominated regions, the H$_2$ fraction reaches a floor value of a few percent, and that star formation within this residual molecular component proceeds on the same timescale as it does in the more H$_2$-rich regions, yielding a ceiling of $\sim 100$ Gyr on the total gas star formation timescale.

Theoretical models to date have been less successful at explaining the behavior of these H~\textsc{i}-dominated regions. While the KMT model successfully predicted the metallicity-dependent location of the transition between the H$_2$-rich and H$_2$-poor regions, and the corresponding dramatic increase in star formation timescale between the two, it did not successfully predict the $\sim 100$ Gyr ceiling on the star formation timescale that appears in the H~\textsc{i}-dominated regime. The alternative model proposed by \citet[hereafter OML]{ostriker10a}, while it did successfully predict a ceiling on the total gas depletion time, failed to reproduce the sharp, metallicity-dependent change in star formation timescale between the H$_2$-rich and H$_2$-poor regimes \citep{bolatto11a}.\footnote{\citet{bolatto11a} proposed a modified version the OML model with an extra metallicity-dependence introduced to fit the observations of the Small Magellanic Cloud. I discuss this model in Section \ref{sec:omlcomparison}.}

The goal of this paper is to extend the KMT model to provide a more accurate treatment of the behavior of H~\textsc{i}-dominated regions. The central idea of this extension is as follows: the way a given section of a galactic disk is partitioned between a non-star-forming H~\textsc{i} phase and the star-forming H$_2$ phase is determined by the gas column density, the metallicity, and the ratio of the ISRF to the density of the cold atomic ISM. If the atomic interstellar medium exists at a pressure where both warm and cold neutral atomic phases are present, then the ratio of ISRF to density is approximately fixed, and the transition becomes a function of the column density and metallicity alone. This is the original KMT model. However, in regions where the depletion time of the gas is as long as 100 Gyr, the ISRF will be extremely small. At sufficiently low ISRF intensity, the ratio of ISRF to density can no longer remain fixed, because the density of the cold atomic gas can only fall so far before its pressure falls below the minimum required to maintain hydrostatic equilibrium. The need to maintain hydrostatic balance sets a floor on the density of the cold atomic phase of the ISM, and we will show that this in turn tends to put a floor on the H$_2$ fraction and the star formation rate. We show that a model including this effect naturally explains both \textit{where} the H$_2$-rich to H$_2$-poor transition occurs as a function of metallicity, and \textit{why} the star formation timescale saturates at $\sim 100$ Gyr in the H~\textsc{i}-dominated region.

The plan for the rest of this paper is as follows. In Section \ref{sec:model} I show how the KMT model can be modified to include the limits imposed by hydrostatic equilibrium at low star formation rate and ISRF strength; I refer to the model that results from this extension as the KMT+ model. Section \ref{sec:observations} contains comparisons between the KMT+ model and a wide variety of observations, both in the local Universe and at high redshift. I discuss some applications of the KMT+ model, and compare to alternative models, in Section \ref{sec:discussion}, and I summarize in Section \ref{sec:summary}.

\section{Model}
\label{sec:model}

\subsection{The Density of Cold Atomic Gas}
\label{sec:ncnm}

Consider a galactic disk in which the atomic interstellar medium consists of a warm phase (WNM) and a cold phase (CNM). \red{I discuss the limits of applicability of this two-phase model in Section \ref{sec:twophaselimits}.} \citet{wolfire03a} show that there is a minimum density for the cold phase of
\begin{equation}
n_{\rm CNM, min} \approx 31 G_0' \frac{Z_d'/Z_g'}{1+3.1(G_0' Z_d'/\zeta_t')^{0.365}}\mbox{ cm}^{-3},
\end{equation}
where $G_0'$ is the intensity of the ISRF, $Z_d'$ and $Z_g'$ are the dust phase and gas phase metallicities, $\zeta_t'$ is the ionization rate due to cosmic rays and X-rays, and primes indicate quantities normalized to their values in the Solar neighborhood. Following KMT, we approximate that $G_0'/\zeta_t'\approx 1$, since both should scale approximately with the local star formation rate, and that $Z_d' = Z_g' = Z'$, since both should scale approximately with the total supply of metals. In two-phase equilibrium, the CNM can exist at a range of densities from $n_{\rm min}$ up to $\sim 5 n_{\rm min}$, and KMT adopt a fiducial value of
\begin{eqnarray}
\label{eq:ncnmkmt}
n_{\rm CNM, 2p} & = & \phi_{\rm CNM} n_{\rm CNM, min} \\
& \approx & 23 G_0' \left(\frac{1+3.1 Z'^{0.365}}{4.1}\right)^{-1}\mbox{ cm}^{-3}
\end{eqnarray}
with $\phi_{\rm CNM} = 3$.

The above expression, taken at face value, would imply that as $G_0'\rightarrow 0$, we should have $n_{\rm CNM}$ and thus the pressure of the CNM approaching 0 as well. However, the pressure of the CNM cannot go to arbitrarily low values, because of the need to maintain hydrostatic balance. Consider a galactic disk consisting of the two atomic phases mentioned above, plus a gravitationally-bound molecular phase that, due to its boundedness, does not contribute to the pressure of the ISM except through its gravity. OML show that the pressure in such a disk may be written as the sum of three components:
\begin{equation}
\label{eq:pmp}
P_{\rm mp} \approx \frac{\pi}{2} G \Sigma_{\rm HI}^2 + \pi G \Sigma_{\rm HI}\Sigma_{\rm H_2} + 2\pi\zeta_d G \frac{\rho_{\rm sd}}{\rho_{\rm mp}} \Sigma_{\rm HI}^2,
\end{equation}
where $\Sigma_{\rm HI}$ and $\Sigma_{\rm H_2}$ are the atomic and molecular gas surface densities, respectively, $\zeta_d\approx 0.33$ is a numerical factor whose exact value depends on the shape of the gas surface density profile, $\rho_{\rm sd}$ is the volume density of stars and dark matter within the gas disk ($\sim 0.01$ $\msun$ pc$^{-3}$ in the Solar neighborhood -- \citealt{holmberg00a}), and $\rho_{\rm mp}$ is the volume-weighted mean gas density as the midplane. Here, the first term in the equation represents the self-gravity of the non-gravitationally-bound H~\textsc{i}, the second term represents the weight of the H~\textsc{i} within the gravitational field provided by the bound H$_2$ clouds, and the third term represents the weight of the H~\textsc{i} within the gravitational field of the stars and dark matter.

OML argue that the thermal pressure at the midplane will be smaller than this by a factor of $\alpha\approx 5$ due to the additional support provided by turbulence, magnetic fields, and cosmic ray pressure, so that $P_{\rm th} = P_{\rm mp}/\alpha$. The thermal pressure in turn can be written
\begin{equation}
\label{eq:pth1}
P_{\rm th} = \rho_{\rm mp} \tilde{f}_w c_w^2,
\end{equation}
where $c_w\approx 8$ km s$^{-1}$ is the sound speed in the WNM, and $\tilde{f}_w$ is the ratio of the mass-weighted mean square thermal velocity dispersion to the square of the warm gas sound speed. The value of $\tilde{f}_w$ is the most uncertain parameter in the model. Following OML, I adopt $\tilde{f}_w = 0.5$ as a fiducial value, but I discuss the basis for this choice, and the implications of a different choice, in Appendix \ref{app:fw}.

Combining equations (\ref{eq:pmp}) and (\ref{eq:pth1}) yields
\begin{eqnarray}
P_{\rm th} & = & \frac{\pi G \Sigma_{\rm HI}^2}{4\alpha} \left\{ 1 + 2 R_{\rm H_2} 
+ \left[\left(1+ 2 R_{\rm H_2}\right)^2\right.\right.
\nonumber \\
& & \left.\left. {}  + \frac{32\zeta_d\alpha \tilde{f}_w c_w^2 \rho_{\rm sd}}{\pi G \Sigma_{\rm HI}^2}\right]^{1/2}\right\},
\label{eq:pth}
\end{eqnarray}
where $R_{\rm H_2} \equiv \Sigma_{\rm H_2}/\Sigma_{\rm HI}$. Note that equation (\ref{eq:pth}) is, except for some changes in notation, identical to equation (11) of OML.

We can then ask: what is the minimum possible density that the CNM can have? \citet{wolfire03a} show that the CNM can only exist up to a maximum temperature $T_{\rm CNM,max} \approx 243$ K, and thus the smallest possible density the CNM can have and still maintain hydrostatic balance is
\begin{equation}
\label{eq:ncnmhd}
n_{\rm CNM,hydro} = \frac{P_{\rm th}}{1.1 k_B T_{\rm CNM,max}},
\end{equation}
where the factor of $1.1$ accounts for He. The central \textit{ansatz} of this work is that the above expression represents a floor on the CNM density, and that the CNM density will therefore be
\begin{equation}
\label{eq:ncnm}
n_{\rm CNM} = \max\left(n_{\rm CNM,2p}, n_{\rm CNM,hydro}\right).
\end{equation} 
\red{Note that the exact numerical value 243 K for $T_{\rm CNM,max}$ is a result of \citeauthor{wolfire03a}'s simplified analytic model, and that the results obtained from numerical calculations can be a factor of $\sim 50\%$ larger or smaller at Solar metallicity, and vary by somewhat larger factors at non-Solar metallicity. As I show below, this uncertainty will translate directly into an uncertainty of comparable magnitude in the predicted star formation rate in the regime where $n_{\rm CNM} = n_{\rm CNM,hydro}$. However, this is probably a smaller effect than the uncertainty in $\tilde{f}_w$.
}

\subsection{The Molecular Gas Fraction}

Given this limit on the CNM density, we can now use the KMT formalism to compute how the gas is partitioned between the atomic and molecular phases. This transition is governed by the total column density, the metallicity, and the dimensionless radiation field parameter
\begin{equation}
\label{eq:chi}
\chi = \frac{f_{\rm diss} \sigma_d c E_0^*}{n_{\rm CNM} \mathcal{R}} = 7.2 \frac{G_0'}{n_1},
\end{equation}
where $f_{\rm diss} \approx 0.1$ is the fraction of absorptions of a Lyman-Werner band photon by an H$_2$ molecule that result in dissociation, $\sigma_d \approx 10^{-21} Z'$ cm$^{-2}$ is the dust absorption cross section per H nucleus for Lyman-Werner band photons, $E_0^* = 7.5\times 10^{-4} G_0'$ cm$^{-3}$ the free-space density of Lyman-Werner band photons, $n_{\rm CNM}$ is the CNM density, $\mathcal{R} \approx 10^{-16.5}Z'$ cm$^3$ s$^{-3}$ is the rate coefficient for H$_2$ formation on dust grains, and $n_1 = n_{\rm CNM}/10$ cm$^{-3}$. \citet{mckee10a} show that the H$_2$ fraction $f_{\rm H_2} \equiv \Sigma_{\rm H_2}/(\Sigma_{\rm HI} + \Sigma_{\rm H_2})$ is well-approximated by
\begin{equation}
\label{eq:fH2}
f_{\rm H_2} = 
\left\{
\begin{array}{ll}
1 - (3/4)s/(1+0.25 s), & s < 2 \\
0, & s \ge 2
\end{array}
\right.,
\end{equation}
where
\begin{eqnarray}
s & \approx & \frac{\ln\left(1 + 0.6 \chi + 0.01\chi^2\right)}{0.6\tau_{c}} \\
\tau_c & = & 0.066 f_c Z' \Sigma_0,
\label{eq:tauc}
\end{eqnarray}
where $\Sigma_0 = \Sigma/1$ $\msun$ pc$^{-2}$, $\Sigma = \Sigma_{\rm HI} + \Sigma_{\rm H_2}$, and $f_c$ is a clumping factor that represents the ratio of the surface densities characteristic of atomic-molecular complexes to the surface density averaged over the scale being observed. On scales of $\sim 100$ pc or less, $f_c$ is simply unity, while on scales of $\sim 1$ kpc, $f_c\approx 5$.

\red{
I note that the KMT formalism is based on the idea that the hydrogen is in chemical equilibrium, and at sufficiently low metallicity this assumption breaks down because the chemical equilibration time becomes long compared to the galaxy dynamical time. \citet{krumholz11a} compare the predictions of the KMT model to fully-time dependent simulations, and find good agreement at metallicities $Z'\ga 0.01$. Analytic models by \citet{krumholz12e} also predict a breakdown of equilibrium between $Z'=0.01$ and $0.1$. In contrast, \citet{mac-low12a} find in their simulations that the H$_2$ abundance is non-equilibrium even at Solar metallicity. One possible explanation for the apparent discrepancy between this result and that of \citeauthor{krumholz11a} is that \citeauthor{mac-low12a} simulated $<20$ pc-length periodic boxes, a region much smaller than a galactic scale height, or even the size of a large GMC. In contrast, \citeauthor{krumholz11a} used simulations with much lower resolution, but that covered an entire galaxy and thus were able to follow galaxy-scale flows and processes, and compared the analytic predictions to numerical results on scales of $65$ pc. If this is indeed the source of the disagreement, then use of the KMT model here is reasonable, as the quantity of interest in what follows is the mean H$_2$ fraction averaged over $>100$ pc scales. Moreover, at least in Solar neighborhood clouds, there is now direct observational evidence that the KMT model correctly predicts observed H~\textsc{i} and H$_2$ column densities, indicating that chemical equilibrium is a reasonable approximation \citep{lee12a}.
}

\subsection{Star Formation}

Equation (\ref{eq:fH2}), using $\chi$ evaluated with $n_{\rm CNM}$ from equation (\ref{eq:ncnm}), provides an estimate of the H$_2$ fraction that properly includes the limits on $n_{\rm CNM}$ imposed by hydrostatic balance. However, we are not yet in a position to evaluate it, because $\chi$ depends on $G_0'/n_{\rm CNM}$. In the case where $n_{\rm CNM} = n_{\rm CNM,2p}$, substituting the value of $n_{\rm CNM,2p}$ from equation (\ref{eq:ncnmkmt}) into equation (\ref{eq:chi}) gives
\begin{equation}
\label{eq:chi2p}
\chi = \chi_{\rm 2p} \equiv 3.1 \left(\frac{1+3.1 Z'^{0.365}}{4.1}\right),
\end{equation}
which depends on metallicity alone. This is original KMT model. To handle the general case, however, we must also solve for $G_0'/n_{\rm CNM}$ in the case of a disk that is at the minimum possible CNM density, $n_{\rm CNM,hydro}$. To do so, I follow OML is approximating that $G_0'$ is proportional to the star formation rate per unit area, with the normalization set by the conditions in the Solar neighborhood:
\begin{equation}
\label{eq:g0}
G_0' \approx \frac{\dot{\Sigma}_*}{\dot{\Sigma}_{*,0}}
\end{equation}
with $\dot{\Sigma}_{*,0} = 2.5\times 10^{-3}$ $\msun$ pc$^{-2}$ Myr$^{-1}$. \red{I discuss the limits of this approximation in Section \ref{sec:twophaselimits}.}

The star formation rate per unit area depends on the H$_2$ abundance and on the properties of the star-forming molecular clouds. \citet{krumholz12a} show that all available observations of star-forming molecular clouds, from the scales of individual clouds in the Milky Way to entire starburst galaxies, are consistent with a universal star formation law
\begin{equation}
\label{eq:sfr}
\dot{\Sigma}_* = f_{\rm H_2}\epsilon_{\rm ff} \frac{\Sigma}{t_{\rm ff}},
\end{equation}
where $\epsilon_{\rm ff} \approx 0.01$ and $t_{\rm ff}$ is the free-fall time of the molecular gas. The value of $\epsilon_{\rm ff}$ may be understood quantitatively as resulting from supersonic turbulence in the GMCs \citep{krumholz05c, padoan11a, hennebelle11b, federrath12a}. In galaxies with very high surface densities, where the entire ISM forms a contiguous molecular medium, the latter quantity is set by the condition that the Toomre $Q$ of the galactic disk be about unity. However, in galaxies like the Milky Way where molecular clouds are discreet self-gravitating entities, \citeauthor{krumholz12a}~show that the free-fall time is well-approximated by
\begin{equation}
\label{eq:tff}
t_{\rm ff} \approx \frac{\pi^{1/4}}{\sqrt{8}} \frac{\sigma_g}{G(\Sigma_{\rm GMC}^3 \Sigma)^{1/4}}
\approx 31\Sigma_0^{-1/4}\mbox{ Myr},
\end{equation}
where $\sigma_g \approx 8$ km s$^{-1}$ is the velocity dispersion of the galactic disk and $\Sigma_{\rm GMC} \approx 85$ $\msun$ pc$^{-2}$ is the characteristic surface density of self-gravitating molecular clouds. Since this is the regime that generally applies when we are near the H~\textsc{i} to H$_2$ transition, we specialize to it, but we caution that equation (\ref{eq:tff}) is only valid for galaxies in the GMC regime, and we refer readers to \citeauthor{krumholz12a}~for a more thorough discussion.

With this specification for $t_{\rm ff}$, equations (\ref{eq:ncnm}), (\ref{eq:chi}), (\ref{eq:fH2}), (\ref{eq:g0}), and (\ref{eq:sfr}) constitute a complete set of equations in the unknowns $n_{\rm CNM}$, $\chi$, $f_{\rm H_2}$, $G_0'$, and $\dot{\Sigma}_*$, which may be solved for any specified combination of total gas surface density $\Sigma$, density of stars and dark matter $\rho_{\rm sd}$, and metallicity $Z'$. In practice numerical solution is more straightforward than would ordinarily be the case for a a system of five non-linear equations, because several of the equations are trivial and as a result the system can be reduced to two a pair of single-variable non-linear equations as follows. Given a specified value of $G_0'$, as well as $\Sigma$, $\rho_{\rm sd}$, and $Z'$, it is trivial to combine equations (\ref{eq:ncnm}), (\ref{eq:chi}), (\ref{eq:fH2}) into a single non-linear equation for $f_{\rm H_2}$. One can then solve iteratively: guess a value for $G_0'$, solve for $f_{\rm H_2}$, and then use $f_{\rm H_2}$ to compute $\dot{\Sigma}_*$ from equation (\ref{eq:sfr}). In general the pair $(G_0', \dot{\Sigma}_*)$ that results from this procedure will not satisfy equation (\ref{eq:g0}), but one may then iterate on $G_0'$ using standard methods (e.g.~Newton's method or Brent's method) to find the pair $(G_0', \dot{\Sigma}_*)$ that does satisfy equation (\ref{eq:g0}).

\begin{figure*}
\begin{center}
\includegraphics[width=6.5in]{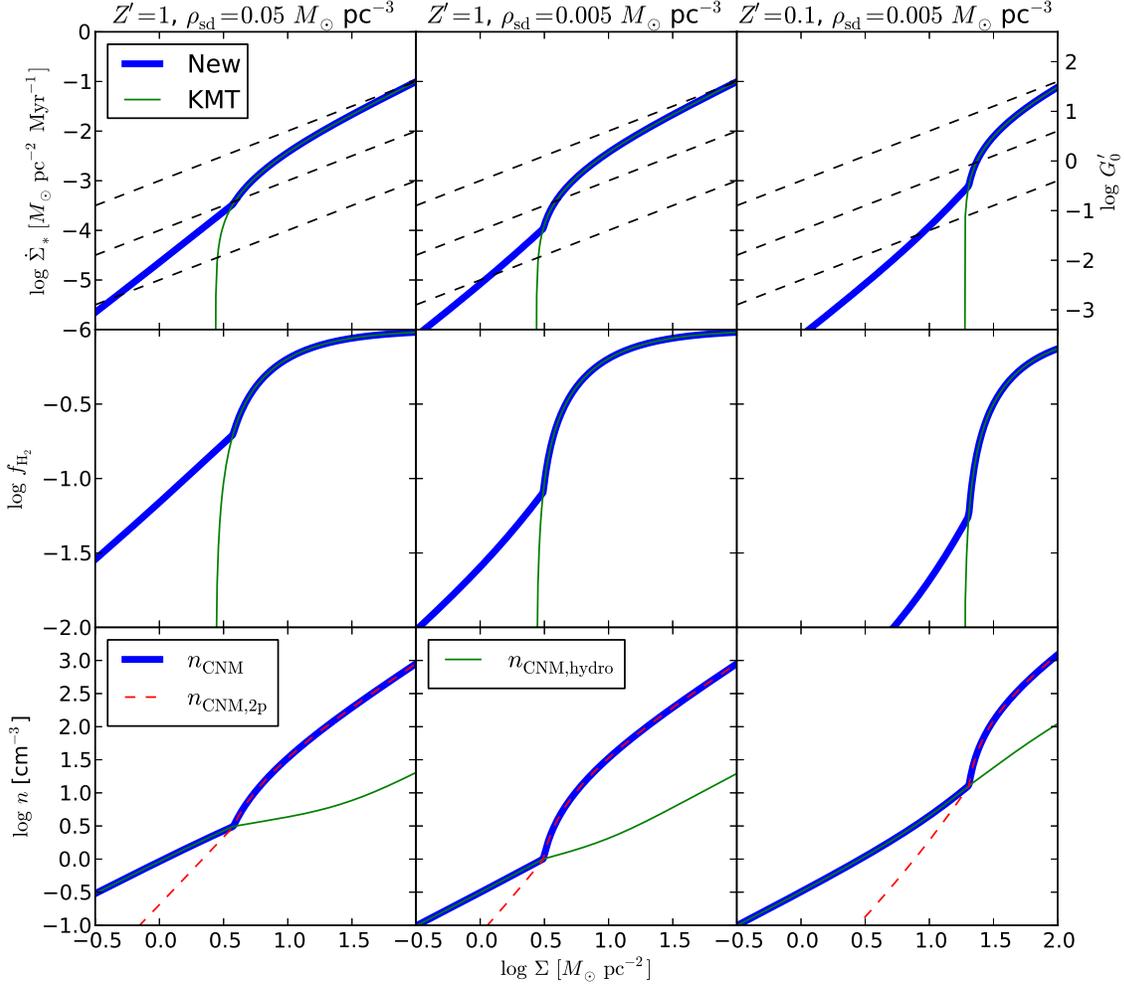}
\end{center}
\caption{
\label{fig:samplesol}
Sample solutions for various quantities as a function of gas surface density $\Sigma$. The columns show, from left to right, the solutions we obtain for $(Z',\rho_{\rm sd})$ values of $(1,0.05)$, $(1, 0.005)$, and $(0.1,0.005)$, as indicated, where $\rho_{\rm sd}$ is in units of $M_\odot$ pc$^{-3}$. The top row shows the star formation rate $\dot{\Sigma}_*$, and the local FUV radiation field $G_0'$. Thick lines show the new model introduced in this paper, while thin lines show the results of the original KMT model. Dashed black lines indicate constant depletion times $t_{\rm dep}\equiv \Sigma/\dot{\Sigma}_*$ of 1, 10, and 100 Gyr, from top to bottom The middle row shows $f_{\rm H_2}$, again for the present model (thick lines) and the original KMT model (thin lines). The bottom row shows the CNM density: the thick solid line shows the density computed from equation (\ref{eq:ncnm}), while the thin solid and dashed lines show $n_{\rm CNM,2p}$ (equation \ref{eq:ncnmkmt}) and $n_{\rm CNM,hydro}$ (equation \ref{eq:ncnmhd}). The former is calculated using the value of $G_0'$ shown in the upper row.
}
\end{figure*}

Figure \ref{fig:samplesol} we show some sample solutions to the system of equations. The qualitative behavior of the results can be understood as follows. Where the column density is high, the H$_2$ fraction is also high, and the star formation rate is relatively high, and the density of the CNM is $n_{\rm CNM,2p}$, the value expected for two-phase equilibrium. As the surface density drops, so do the star formation rate, H$_2$ fraction, the ISRF intensity, and the two-phase CNM density. Once the two-phase CNM density drops below the minimum allowed by hydrostatic equilibrium, the star formation rate-column density relation breaks, as the density continues to drop but following $n_{\rm CNM,hydro}$ instead of $n_{\rm CNM, 2p}$. The location of the break, and the track that the star formation rate and H$_2$ fraction follow below it, depend on $\rho_{\rm sd}$, since this influences the thermal pressure and thus $n_{\rm CNM,2p}$.

One subtlety worth pointing out is that, while both $n_{\rm CNM,2p}$ and $n_{\rm CNM,hydro}$ rise with surface density, the former does so faster than the latter, which is why galaxies tend to bump up against the CNM volume density floor when their surface densities are low, not high. The minimum volume density required by hydrostatic equilibrium, $n_{\rm CNM,hydro}$, scales with surface density to a power between 1 and 2, depending on which term in equation (\ref{eq:pmp}) dominates. In contrast, the two-phase equilibrium density $n_{\rm CNM,2p}$ is proportional to the star formation rate, which varies as a much higher power of the total gas column density in the column density range around the H~\textsc{i} - H$_2$ transition. Thus the fact that galaxies are near their volume density floors at small rather than large surface densities is a direct result of the steep non-linearity of the star formation as a function of surface density when one is near the threshold where the ISM is transitioning from H$_2$-rich to H$_2$-poor.

\subsection{Limiting Behaviors and Analytic Approximation}
\label{sec:limits}

While it is straightforward to solve the equations numerically, we can gain additional physical insight by developing analytic approximations. We begin by examining the behavior in the H$_2$-poor and H$_2$-rich limits, starting with the former. The defining feature in this regime is that $f_{\rm H_2} \ll 1$, and so to obtain an approximate analytic solution we can linearize the equations to first order around $f_{\rm H_2} = 0$. Following this procedure, equations (\ref{eq:chi}) -- (\ref{eq:g0}) reduce to
\begin{equation}
f_{\rm H_2} \approx \frac{1}{3}\left(2 - \frac{44\mbox{ Gyr}}{f_c Z' n_1}\frac{\dot{\Sigma}_*}{\Sigma}\right).
\end{equation}
Using this in equation (\ref{eq:sfr}) and re-arranging, we obtain
\begin{equation}
\label{eq:tdeplim}
t_{\rm dep} \equiv \frac{\Sigma}{\dot{\Sigma}_*} \approx \frac{3t_{\rm ff}}{2\epsilon_{\rm ff}} + \frac{22\mbox{ Gyr}}{f_c Z' n_1}.
\end{equation}
In the regime of low star formation rate, the first of these terms, which is of order 2 Gyr, is generally smaller than the second, and thus to good approximation the depletion time simply scales as the inverse of the CNM volume density, which in this limit is $n_{\rm CNM,hydro}$ rather than $n_{\rm CNM,2p}$. In the limit $f_{\rm H_2}\ll 1$, equations (\ref{eq:pth}) and (\ref{eq:ncnmhd}) reduce to
\begin{eqnarray}
n_{\rm CNM,hydro} & \approx &  \frac{\pi G \Sigma^2}{4\alpha (1.1 k_B T_{\rm CNM,max})} 
\nonumber \\
& &
{} \cdot \left[1+\left(1+\frac{32\zeta_d\alpha \tilde{f}_w c_w^2 \rho_{\rm sd}}{\pi G \Sigma^2}\right)^{1/2}\right].
\label{eq:ncnmhdapprox}
\end{eqnarray}
The final term, which comes from stellar and dark matter gravity, is usually much greater than unity at small $\Sigma_{\rm HI}$ and reasonable values of $\rho_{\rm sd}$, and thus we can drop the factors of unity to obtain
\begin{equation}
n_{\rm CNM,hydro} \approx \sqrt{\frac{2\pi G \zeta_d \tilde{f}_w \rho_{\rm sd}}{\alpha}}\left(\frac{c_w}{1.1 k_B T_{\rm CNM,max}}\right) \Sigma
\end{equation}
Substituting this into equation (\ref{eq:tdeplim}), we obtain
\begin{equation}
\label{eq:tdephdstar}
t_{\rm dep,hd,*} \approx \frac{3.1\mbox{ Gyr}}{\Sigma_0^{1/4}} + \frac{100\mbox{ Gyr}}{(f_c/5) Z' \rho_{\rm sd,-2}^{1/2} \Sigma_0}
\end{equation}
where $\rho_{\rm sd,-2} = \rho_{\rm sd}/0.01$ $\msun$ pc$^{-3}$. In the case where the term proportional to $\rho_{\rm sd}$ in equation (\ref{eq:ncnmhdapprox}) is much smaller than unity, we can drop it, and the same procedure yields
\begin{equation}
\label{eq:tdephdgas}
t_{\rm dep,hd,gas} \approx \frac{3.1\mbox{ Gyr}}{\Sigma_0^{1/4}} + \frac{360\mbox{ Gyr}}{(f_c/5) Z' \Sigma_0^2}.
\end{equation}
We therefore have arrived at a quantitative explanation for why galaxies should show $\sim 100$ Gyr depletion times in the H$_2$-poor regime. 

We can take a similar approach in the H$_2$-rich, high surface density regime by expanding to first order about $1-f_{\rm H_2}$. Note that $1-f_{\rm H_2} \ll 1$ corresponds to the case where $s \ll 1$ in equation (\ref{eq:fH2}), and thus
\begin{equation}
1 - f_{\rm H_2} \approx \frac{3}{4}s = 1.25 \frac{\ln(1+0.6\chi+0.01\chi^2)}{\tau_c}.
\end{equation}
In this case we have
\begin{eqnarray}
\Sigma_{\rm HI,max} & = & \left(1-f_{\rm H_2}\right)\Sigma 
\\
& \approx & \frac{19}{f_c Z'} \ln(1+0.6\chi+0.01\chi^2)\,\msun\mbox{ pc}^{-2}
\\
& \approx & \frac{24}{f_c Z'} \left[\frac{\ln(1+0.6\chi+0.01\chi^2)}{1.29}\right] \,\msun\mbox{ pc}^{-2}.
\label{eq:himax}
\end{eqnarray}
In the last step we have used the fact that, in the H$_2$-rich regime, we expect $\chi$ to approach the value expected for two-phase equilibrium, $\chi = 3.1 \left(1+3.1 Z'^{0.365}\right)/4.1$, and the normalization factor in the final term is the chosen to that the term in parenthesis is unity for $Z' = 1$. The above equation constitutes a prediction of the maximum possible H~\textsc{i} column; higher total columns result in the excess gas taking the form of H$_2$ rather than H~\textsc{i}. Finally, the depletion time is simply
\begin{equation}
\label{eq:tdep2p}
t_{\rm dep} = \frac{t_{\rm ff}}{f_{\rm H_2}\epsilon_{\rm ff}} = \frac{3.1\mbox{ Gyr}}{f_{\rm H_2}\Sigma_0^{1/4}}.
\end{equation}
where $f_{\rm H_2}$ is evaluated from equation (\ref{eq:fH2}) using $\chi = \chi_{\rm 2p}$ (equation \ref{eq:chi2p}). Note that this is nearly the same as equation (10) of \citet{krumholz09b}. The exponent is slightly different because the free-fall time has been estimated slightly differently, but the actual numerical value of $t_{\rm dep}$ differ by at most a factor of $\sim 2$ over the full range from $\sim 10 - 100$ $\msun$ pc$^{-2}$ where this equation applies. Similarly, over this range $t_{\rm dep}$ varies by less than a factor of 2 from the constant value of 2 Gyr adopted by OML.

\begin{figure}
\begin{center}
\includegraphics[width=3.5in]{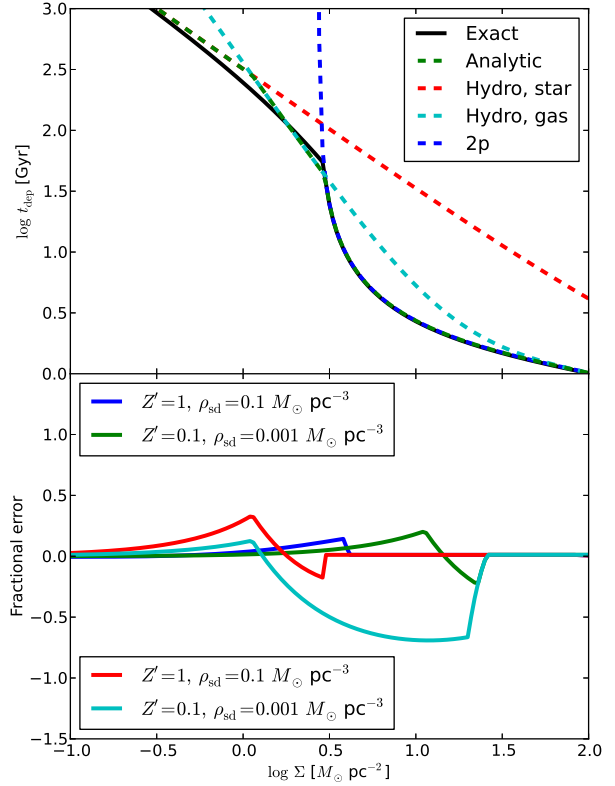}
\end{center}
\caption{
\label{fig:kmtapprox}
Comparison between the exact numerical solution to our model and an analytic approximation. The top panel shows the case $Z'=1$, $\rho_{\rm sd} = 0.001$ $\msun$ pc$^{-3}$. The black solid line gives the exact value of $t_{\rm dep}$ that results from a numerical solution to the equations, the green dashed line is the analytic approximation given by equation (\ref{eq:tdepanalyt}), and the red, cyan, and blue dashed lines are the approximate values of $t_{\rm dep}$ given by equations (\ref{eq:tdephdstar}), (\ref{eq:tdephdgas}), and (\ref{eq:tdep2p}); these correspond to the H$_2$-poor, stellar-dominated regime, the H$_2$-poor, gas-dominated regime, and the H$_2$-rich regimes, respectively. The bottom panel shows the fractional error in the analytic approximation, defined as $(t_{\rm dep,analyt}-t_{\rm dep,exact})/t_{\rm dep,exact}$, for the four cases of metallicity and stellar density listed in the legend. All models use $f_c = 5$.
}
\end{figure}

The total gas depletion time is roughly the minimum of the three depletion timescales computed above, i.e.
\begin{equation}
\label{eq:tdepanalyt}
t_{\rm dep} \approx \min(t_{\rm dep,2p}, t_{\rm dep,hd,*}, t_{\rm dep,hd,gas}).
\end{equation}
The first case is the molecule-rich one treated by the original KMT model, the middle one is the case of a molecule-poor galaxy where the pressure of the gas is dominated by the gravity of the stellar or dark matter component, and the final case corresponds to a molecule-poor galaxy where gas self-gravity dominates. Figure \ref{fig:kmtapprox} shows how this approximation compares to the exact numerical solution. As the plot shows, the approximate value of $t_{\rm dep}$ matches the exact numerical result to better than a factor of 2 over the range $Z'=0.1 - 1$, $\rho_{\rm sd} =0.001 - 0.1$ $\msun$ pc$^{-3}$, and is generally more accurate than that. If for a given application it is desirable that $t_{\rm dep}$ be continuous, equation (\ref{eq:tdepanalyt}) could be replaced by a harmonic or squared harmonic mean of the three terms instead of a simple minimum. This is only very marginally less accurate.

\subsection{\red{Domain of Applicability}}
\label{sec:twophaselimits}

\red{
The theory I present here relies on a number of assumptions about the workings of the ISM, and these will apply over a limited range of metallicity and star formation rate. Here I discuss the limitations imposed by those assumptions. First, this model relies on the existence of a distinct CNM phase, but this in turn requires that at least the cool gas be able to reach thermal equilibrium. (It is \textit{not} necessary for this theory at there be a distinct WNM phase in thermal equilibrium, at least in the regime where the hydrostatic pressure is dominant.) The requirement that CNM gas be able to reach thermal equilibrium quickly is satisfied at Solar metallicity \citep[e.g.,][]{wolfire03a}, but at lower metallicity radiative cooling times are longer, and thus one expects that, at sufficiently low metallicity, the model presented here must break down. This is of concern because many of the H~\textsc{i}-dominated regions to which this model is intended to be applied have noticeably sub-Solar metallicities. Nearby spiral galaxies typically have metal gradients of $\sim -0.03$ dec kpc$^{-1}$, \citep[e.g.,][]{vila-costas92a, considere00a, pilyugin04a, kennicutt11a}, so their H~\textsc{i}-dominated regions at $\sim 15$ kpc from the galactic center typically have metallicities of $Z' \sim 0.5$. Similarly, the mass-metallicity relation \citep[e.g.,][]{tremonti04a} implies that dwarf galaxies generally have sub-Solar metallicities, with a typical value of $Z'\sim 0.3$ at a stellar mass of $\sim 10^9$ $M_\odot$. High-redshift systems such as DLAs may have even lower metallicities, $Z'\sim 0.01-0.1$ \citep{prochaska03a, rafelski12a}. Thus one must ask whether the model I present here can reasonably be applied to these systems.
}

\red{
To estimate whether the gas should be close to or far from thermal equilibrium, one must compare the thermal equilibration time to some mechanical or dynamical timescale over which mechanical forces will change the gas density or temperature. The thermal time is simply the time for a gas at some out-of-equilibrium temperature $T$ to return to thermal equilibrium, and may be formally written
\begin{equation}
t_{\rm therm} = \frac{k_B T}{\Lambda},
\end{equation}
where $\Lambda$ is the gas cooling rate per H atom. Following \citet{krumholz12e}, if the gas is dominated by C~\textsc{ii} cooling, then the cooling rate in gas of density $n$ is
\begin{equation}
\Lambda_{\rm CII} = k_{\rm CII-H} \delta_C k_B T_{\rm CII} \mathcal{C} n,
\end{equation}
where $k_{\rm CII-H} \approx 8\times 10^{-10} e^{-T_{\rm CII}/T}$ cm$^{-3}$ s$^{-1}$ is the rate coefficient for collisional excitation of C~\textsc{ii} by H \citep{launay77a, barinovs05a, schoier05a}, $T_{\rm CII} = 91$ K is the energy of the radiating C~\textsc{ii} level divided by $k_B$, $\delta_C \approx 1.1\times 10^{-4} Z'$ is the carbon abundance relative to H \citep{draine11a}, and $\mathcal{C} \equiv \langle n^2\rangle/\langle n\rangle^2 > 1$ is the clumping factor that represents the increase in the rate of collisional processes due to density inhomogeneity. The choice of dynamical time to which this should be compared is less obvious. One option is simply the free-fall time, which will describe the rate at which self-gravity can alter gas properties. \citet{krumholz12e} shows that the ratio of thermal and free-fall timescales is
\begin{equation}
\frac{t_{\rm therm}}{t_{\rm ff}} = 2.6\times 10^{-4} \left(\frac{T}{T_{\rm CII}}\right) e^{T_{\rm CII}/T} Z'^{-1} \mathcal{C}_{-1}^{-1} n_1^{-1/2},
\end{equation}
where $\mathcal{C}_1 = \mathcal{C}/10$. From this expression it is clear that even gas that is transiently heated to $T \sim 1000$ K can cool to its equilibrium temperature is much less than a free-fall time unless the metallicity is extremely small, $Z' \la 0.01$.
}

\red{
Alternatively, one might to take the mechanical timescale to be the time between shocks from the turbulence in the gas, since these can induce transient heating. Following \citet{wolfire03a}, in a turbulent medium the time between shocks capable of altering the gas temperature by a factor of $\sim 2$ is of order
\begin{equation}
t_{\rm shock} \sim \frac{\lambda_s}{\sigma_{\rm th}},
\end{equation}
where $\lambda_s$ is the sonic length, defined as the length scale for which the thermal and non-thermal velocity dispersions are comparable, and $\sigma_{\rm th}$ is the pre-shock thermal velocity dispersion. To compute $\lambda_s$, we can let $\sigma$ be the non-thermal velocity dispersion at the outer scale of the turbulence $H$, presumably comparable to the galactic scale height, and adopt the usual Burgers' turbulence linewidth-size relation $\sigma(\ell) \propto \ell^{1/2}$ \citep[e.g.,][]{mckee07a}. With this scaling, we have $\lambda_s \approx (\sigma_{\rm th}/\sigma)^2 H$, and the ratio of the shock and thermal timescale becomes
\begin{equation}
\frac{t_{\rm therm}}{t_{\rm shock}} \approx 0.07 \left(\frac{T}{T_{\rm CII}}\right) e^{T_{\rm CII}/T} Z'^{-1} \mathcal{C}_{-1}^{-1} \left(\frac{\sigma}{7\mbox{ km s}^{-1}}\right)^2 \Sigma_0^{-1},
\end{equation}
where in the numerical evaluation I have used $\sigma_{\rm th}^2 = k_B T_{\rm CNM,max}/\mu m_{\rm H}$ with $\mu=1.4$, appropriate for CNM gas at the fiducial temperature in the model, and I have set $\Sigma = n H \mu m_{\rm H}$. The fiducial value of $\sigma = 6$ km s$^{-1}$ to which I have scaled is a typical observed velocity dispersion in regions of very low surface density and star formation rate \citep{stilp13a}; note that this is probably an upper limit, because the velocity dispersion $\sigma$ that should enter this expression is the non-thermal velocity dispersion in the cold gas, which is necessarily smaller. The implication of this conclusion is that the existence of CNM gas in thermal equilibrium is probably a reasonable assumption down to metallicities of $Z'\sim 0.1$, and possibly lower, depending on the strength of the non-thermal motions in the cold gas.
}

\red{
The final assumption that limits the domain of applicability of this model is that the FUV radiation field, as parameterized by $G_0'$, is dominated by local star formation, so that $G_0' \propto \dot{\Sigma}_*$. At a sufficiently low star formation rate, this assumption must break down, since there is non-zero FUV extragalactic background. Following \citet{wolfire03a}, I adopt the parameterization of \citet{draine78a} to describe the radiation field in the Solar neighborhood, and assume that this spectral shape is invariant as the star formation rate varies. I compare this to a diffuse background field taken from the model of \citet{haardt12a}.}

\red{The comparison requires some care, as the spectral shapes are quite different, and thus the ratio of energy densities depends on the energy range over which the local and background radiation fields are compared. One possible choice is $8-13.6$ eV, the energy range that dominates grain photoelectric heating \citep{draine11a}, while another is $11-13.6$ eV, the energy range that dominates H$_2$ photodissociation. With the former choice, I find that the energy density in the FUV background at $z=0$ is smaller than the Solar neighborhood radiation field by a factor of $500$, and, using equation (\ref{eq:g0}), this implies that the locally-produced radiation field dominates for star formation rates $\dot{\Sigma}_*>5.1\times 10^{-6}$ $\msun$ pc$^{-2}$ Myr$^{-1}$. Using the latter choice, the FUV background is weaker by a factor of $1700$, and local star formation dominates as long as $\dot{\Sigma}_*>1.5\times 10^{-6}$ $\msun$ pc$^{-2}$ Myr$^{-1}$. The background FUV radiation field reaches its maximum intensity at $z=(3.2, 3.4)$, where it is weaker than the Solar neighborhood field by a factor of $(3.0, 7.3)$, giving limiting star formation rates of $\dot{\Sigma}_* = (8.4\times 10^{-4}, 3.4\times 10^{-4})$ $\msun$ pc$^{-2}$ Myr$^{-1}$; here the first number corresponds to the results if one compared the radiation fields over the range $8-13.6$ eV, and the latter to comparing them over the range $11-13.6$ eV. At star formation rates below these limiting values, the theory should be modified by setting $G_0'$ to a constant value rather than scaling it by the star formation rate.
}

\section{Comparison to Observations}
\label{sec:observations}

This section presents comparisons between the KMT+ model described in the previous section and a variety of observations of both the local and high redshift Universe.

\subsection{Local Group Galaxies Resolved at $\sim 1$ kpc Scales}

\begin{figure*}
\begin{center}
\includegraphics[width=6.0in]{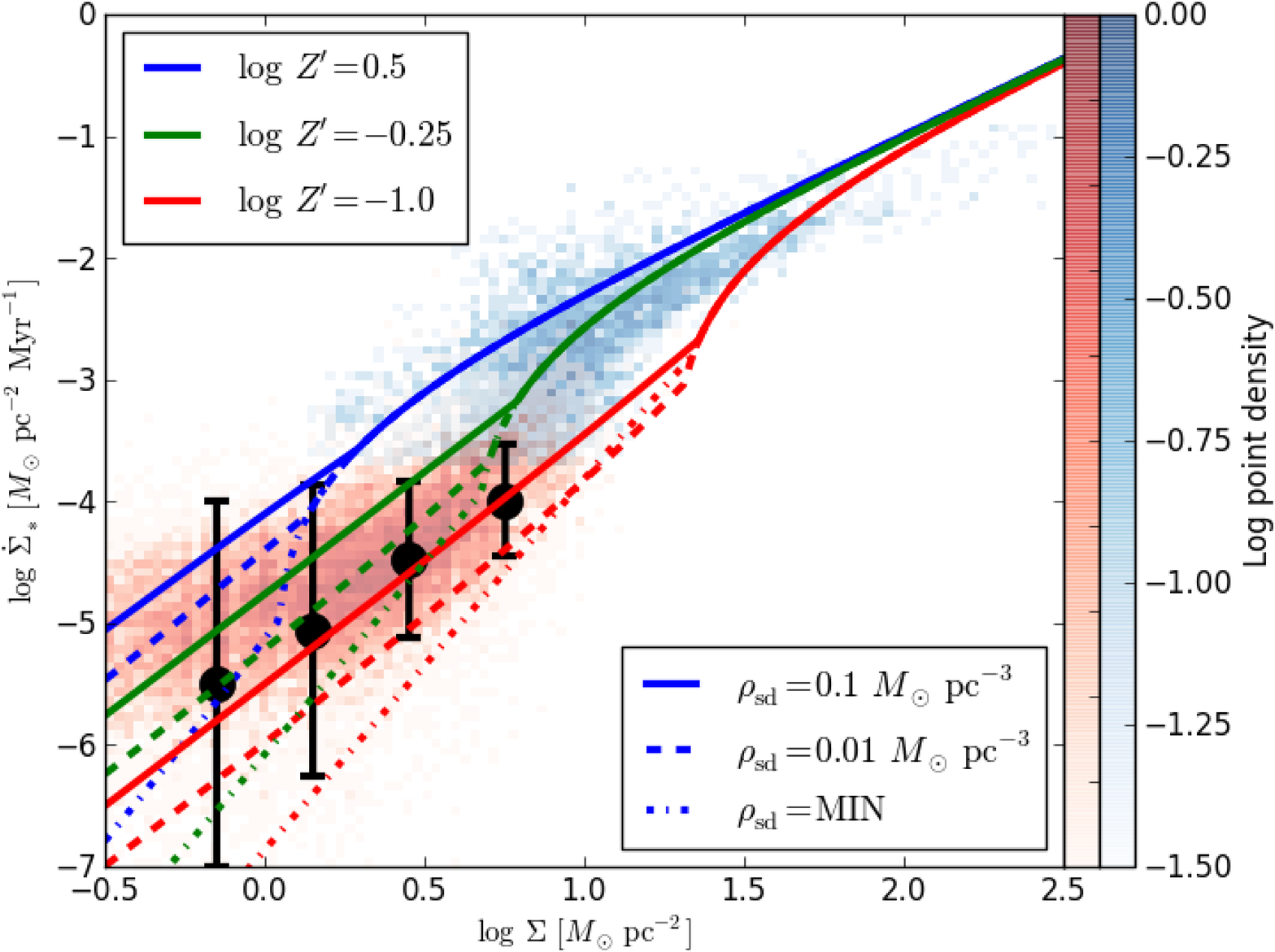}
\end{center}
\caption{
\label{fig:bigielcomp}
Comparison between the model and the observations of \citet{bigiel10a}. Solid lines show the model described in this paper computed with metallicities of $\log Z'=0.5$, $-0.25$, and $-1.0$, as indicated. Solid lines are computed for a star and dark matter density $\rho_{\rm sd} = 0.1$ $\msun$ pc$^{-3}$, dashed lines for $\rho_{\rm sd} = 0.1$ $\msun$ pc$^{-3}$, and dot-dashed lines for the minimum possible value of $\rho_{\rm sd}$, given by equation (\ref{eq:rhosdmin}). The background colors indicate the density of observed data points in the $(\Sigma, \dot{\Sigma}_*)$ as measured for nearby galaxies by \citet{bigiel08a} and \citet{bigiel10a}, normalized so that the pixel with the highest density of points has a value of unity, and with each galaxy in the outer disk region given equal weight. Blue points show the portions of galaxies within $R_{25}$, while red shows the portions outside $R_{25}$. Black points with error bars show the median value and scatter in bins of $\Sigma$ in the outer region. These points are computed by properly accounting for observational errors that produce negative values of $\dot{\Sigma}_*$. These negative values are masked in the logarithmic plot, which is why the red points taper off at low $\dot{\Sigma}_*$. In this range, the black points, which include a correct treatment of errors, should be taken as definitive. See \citet{bigiel10a} for details.
}
\end{figure*}

The first comparison data set is a sample of nearby spiral and dwarf galaxies imaged at resolutions of $\sim 1$ kpc by \citet{bigiel08a, bigiel10a}. This data set includes H~\textsc{i} measurements from 21 cm emission, H$_2$ measurements from CO $J=2\rightarrow 1$ (assuming a fixed CO to H$_2$ conversion factor), and star formation rate measurements from FUV, H$\alpha$, and 24 $\mu$m emission. The data are broken into inner galaxy (inside $R_{25}$; \citealt{bigiel08a}) and outer galaxy (outside $R_{25}$; \citealt{bigiel10a}) parts. The galaxies in the sample span a range of metallicities from $\log Z' = -1.0 - 0.5$.

To compare the KMT+ model to these data, I evaluate the star formation rate predicted by the KMT+ model over the same metallicity range, using a clumping factor $f_c = 5$, following \citet{krumholz09b}, because the data are measured at $\sim 1$ kpc scales. Unfortunately values of $\rho_{\rm sd}$ are not available for most of these galaxies, so one must adopt a reasonable range for the comparison. Inner galaxies may have $\rho_{\rm sd}$ as high as $\sim 1$ $\msun$ pc$^{-3}$, while the Solar neighborhood has $\rho_{\rm sd} \sim 0.01$ $\msun$ pc$^{-3}$ \citep{holmberg00a}. Less is known about how $\rho_{\rm sd}$ falls off in far outer galaxies, but we can obtain an absolute lower limit by considering a galaxy with no stars at all in its outer regions, only dark matter. For a flat rotation curve of speed $V$ produced only by dark matter, $\rho_{\rm sd} = (V/R)^2/(4\pi G)$, where $R$ is the galactocentric radius. The Toomre $Q$ parameter for the gas is
\begin{equation}
Q_g = \frac{\sqrt{2} (V/R) \sigma_g}{\pi G \Sigma},
\end{equation}
so for a star-free galaxy we have
\begin{equation}
\label{eq:rhosdmin}
\rho_{\rm sd} \geq \frac{\pi G Q_g^2 \Sigma^2}{8 \sigma_g^2} = 2.6\times 10^{-5} Q_g^2 \Sigma_0^2\,\msun\mbox{ pc}^{-3}
\end{equation}
where in the numerical evaluation I have used $\sigma_g = 8$ km s$^{-1}$. \citet{elmegreen11a} shows that disks with $Q < 2-3$ will be strongly unstable, and for a star-free disk $Q_g = Q$; the stability threshold differs from the canonical value $Q=1$ because gas, unlike stars, is dissipational and therefore capable of becoming unstable on arbitrarily small scales. Thus  one may obtain a reasonable lower limit on $\rho_{\rm sd}$ by using $Q_g = 2$ in the above question.

In Figure \ref{fig:bigielcomp}, I show the KMT+ model overplotted on the observations. \citet{krumholz09b} have already shown that the KMT model does an excellent job of reproducing the inner galaxy data, and Figure \ref{fig:bigielcomp} shows that the KMT+ model presented here does an excellent job of reproducing the full data set. The original KMT model correctly captured the turn-down in SFR at gas surface densities of $\sim 3-10$ $\msun$ pc$^{-2}$, and the new version also recovers the flattening of $\dot{\Sigma}_*$ versus $\Sigma$ below the turn-down. The model also explains why the observed scatter in SFR at fixed $\Sigma$ is much larger at low surface density than at high surface density. At high surface density, the ISM becomes molecule-dominated, and the model curves for different metallicity and stellar density converge, reducing the scatter. As one approaches the H~\textsc{i}-dominated regime, on the other hand, the metallicity and stellar density both begin to matter a great deal, and a spread in those parameters in the observed galaxy sample leads to a larger scatter in the data. \red{Finally, a caveat is in order: at the lowest star formation rates shown in the Figure, the data are somewhat outside the range of the model's validity, $\dot{\Sigma}_* \ga 3\times 10^{-6}$ $\msun$ pc$^{-2}$ Myr$^{-1}$ (see Section \ref{sec:twophaselimits}), because the star formation rates per unit area are so low that the diffuse UV background should dominate rather than local sources. However, the bulk of the data are in the regime where the model is valid.}

\begin{figure}
\begin{center}
\includegraphics[width=3.5in]{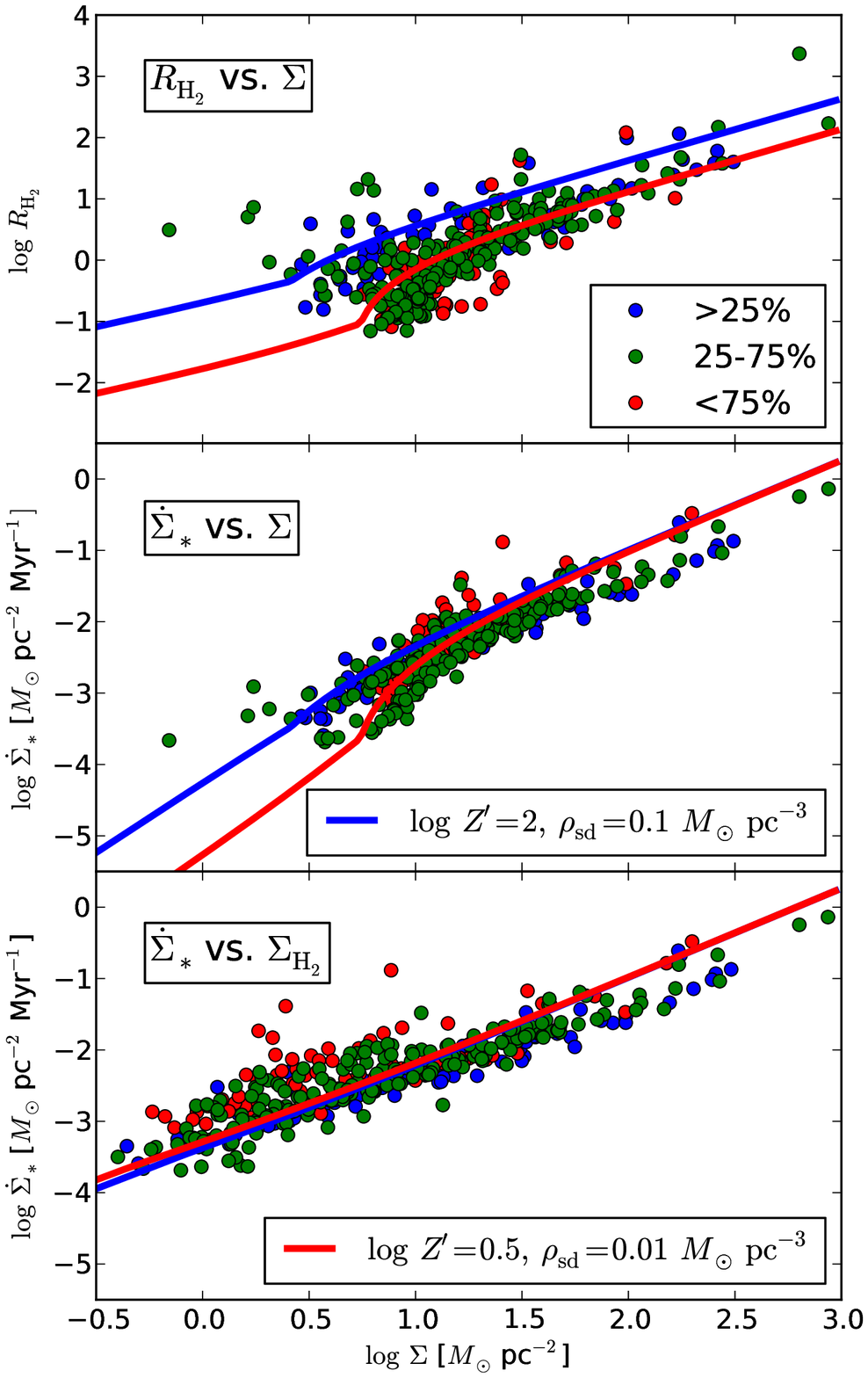}
\end{center}
\caption{
\label{fig:schrubacomp}
Comparison between the model and the observations of \citet{schruba11a}. The top panel shows $R_{\rm H_2}$ versus $\Sigma$, the middle panel shows $\dot{\Sigma}_*$ versus $\Sigma$, and the bottom panel shows $\dot{\Sigma}_*$ versus $\Sigma_{\rm H_2}$. In each panel, circles show azimuthal ring measurements from \citet{schruba11a}, with color indicating metallicity. Blue indicates the top quartile of the sample by metallicity ($12+\log({\rm O}/{\rm H})>8.83$), green indicates middle two quartiles ($8.66 < 12+\log({\rm O}/{\rm H}) < 8.83$), and red indicates bottom quartile ($12+\log({\rm O}/{\rm H})<8.66$). We show only \citeauthor{schruba11a}'s quality 1 and 2 data. Lines show the model described in this paper computed with $Z'=2.0$, $\rho_{\rm sd}=0.1$ $\msun$ pc$^{-3}$ (blue) and $Z' = 0.5$, $\rho_{\rm sd}=0.01$ $\msun$ pc$^{-3}$ (red). For details on how metallicities for the data are determined, see main text.
}
\end{figure}

\subsection{Azimuthal Rings in Local Group Galaxies}

The \citet{bigiel10a} data on outer disks does not contain any information on molecular gas, because in individual $\sim 1$ kpc pixels the CO emission is undetectably small. The quantity reported on the $x$-axis of Figure \ref{fig:bigielcomp} for the outer galaxies is simply the H~\textsc{i} surface density, since the upper limits on molecular surface density imply that it is generally small in comparison. \citet{schruba11a} are able to obtain detections of CO in some of these outer disk regions by stacking the data in radial rings, and this provides us with a second data set to which we can compare in somewhat more detail, since it contains independent data on H~\textsc{i} and H$_2$.

Figure \ref{fig:schrubacomp} shows a comparison between the KMT+ model (again computed with $f_c=5$) and the measurements of \citet{schruba11a}; the three panels show $R_{\rm H_2}$ versus $\Sigma$, $\dot{\Sigma}_*$ versus $\Sigma$, and $\dot{\Sigma}_*$ versus $\Sigma_{\rm H_2}$. As the Figure makes clear, the model does a good job of reproducing all three of these trends. It properly captures the downturn in $R_{\rm H_2}$ and $\dot{\Sigma}_*$ as $\Sigma$ decreases, while at the same time capturing the lack of a corresponding break or curve in the $\dot{\Sigma}_*$ versus $\Sigma_{\rm H_2}$ relation.

Moreover, these data provide an opportunity to begin checking the metallicity-dependence of our model. For each of \citeauthor{schruba11a}'s galaxies, I have obtained a measurement of $\log({\rm O}/{\rm H})$, and I have binned the data into quartiles by $\log({\rm O}/{\rm H})$ value; metallicity data come from, in order of preference, Table 2 of \citet{schruba12a}, Table 7 of \citet{moustakas10a} (using the average of their two calibrations, and taking the radial strip rather than the nuclear values), and Table 1 of \citet{walter08a}. The differences by metallicity are most apparent in the plot of $R_{\rm H_2}$ versus $\Sigma$, where we see that, on average, the highest metallicity quartile has higher $R_{\rm H_2}$ at fixed $\Sigma$, while the lowest metallicity quartile has lower $R_{\rm H_2}$; typical differences in $R_{\rm H_2}$ at fixed $\Sigma$ from the lowest to the highest quartile are $0.2 - 0.6$ dex. This is qualitatively consistent with the predictions of the model.

We should treat this metallicity comparison with caution, since there are a number of potential concerns. First, the H$_2$ abundances in \citet{schruba11a} have been computed using a fixed $\alpha_{\rm CO}$ factor, while in fact we might expect $\alpha_{\rm CO}$ to be lower at lower metallicity \citep[and references therein]{bolatto13a}. The range in metallicity covered by the sample is fairly small ($0.66$ dex), and the lowest metallicity galaxy in the sample as $12+\log({\rm O}/{\rm H})=8.34$, so we might not expect this to be a huge effect; nonetheless, it could potentially explain part of the observed behavior. Second, the metallicities used in constructing the plot represent single values for each galaxy, and do not take into account metallicity gradients, which may vary from galaxy to galaxy, and which will be particularly important at galaxy edges where the metallicity sensitivity is greatest. Third, although were possible I have used oxygen abundance measurements all calibrated on the same scale, uniform calibrations are not available for all the galaxies in the sample, and this may well introduce significant scatter \citep{kewley08a}. Given this caveats, the best that can be said is that the metallicity-dependence that appears in the data is qualitatively consistent with the predictions of the model. In the next few sections, we will examine data that span a much wider range of metallicity, and provide a much more robust test of the metallicity-dependence of KMT+.

\subsection{The Small Magellanic Cloud}

The next comparison data set is the Small Magellanic Cloud (SMC) observations of \citet{bolatto11a}, who measured H~\textsc{i} from 21 cm emission, H$_2$ using dust continuum emission, and star formation using H$\alpha$ plus 24 $\mu$m emission. This data set is a particularly useful test of the model for a number of reasons. First, the galaxy is resolved to better than $100$ pc scales, so there is no need to adopt a clumping factor to account for unresolved structures (i.e.~I set $f_c=1$ throughout this section). This eliminates a free parameter in the model. Second, because the H$_2$ is traced by a means that is independent of an adopted $\alpha_{\rm CO}$, this comparison is not confounded by degeneracy between variations in $\alpha_{\rm CO}$ and real variations in the H$_2$ fraction. Third, the metallicity of the SMC is $Z'\approx 0.2$, providing a large baseline to test the metallicity-dependence predicted by the model.

\begin{figure}
\begin{center}
\includegraphics[width=3.5in]{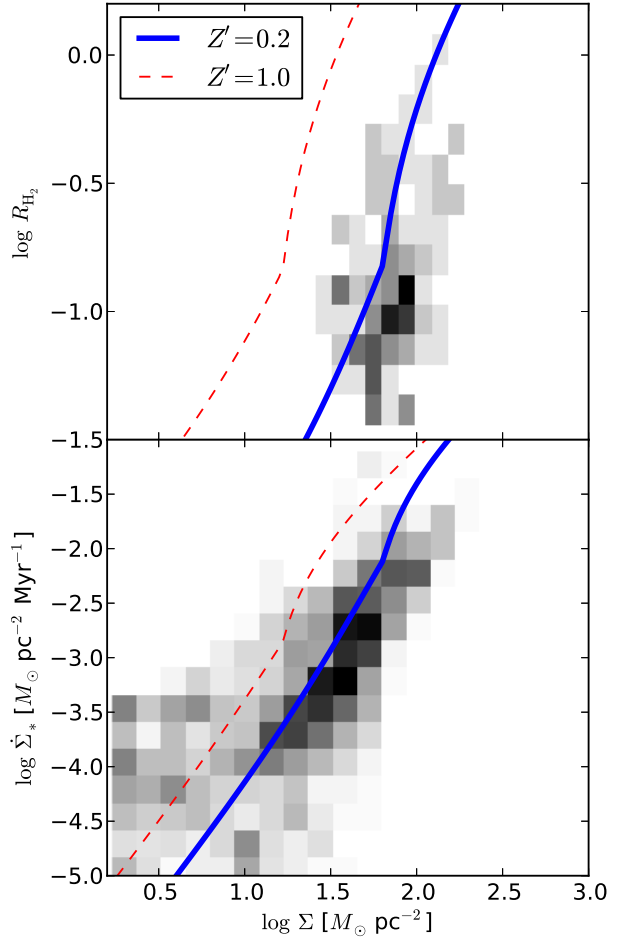}
\end{center}
\caption{
\label{fig:bolattocomp}
Comparison between the KMT+ model and the observations of \citet{bolatto11a}. The top panel shows $R_{\rm H_2}$ versus $\Sigma$, while the bottom panel shows $\dot{\Sigma}_*$ versus $\Sigma$. In each panel, the black and white raster plot shows the density of SMC lines of sight in the $(\Sigma, R_{\rm H_2})$ and $(\Sigma, \dot{\Sigma}_*)$ planes, respectively, with the intensity of the color from white to black proportional to the number of points in each bin. Blue lines show the KMT+ model computed with $Z'=0.2$, $\rho_{\rm sd}=0.02$ $\msun$ pc$^{-3}$; the value of $\rho_{\rm sd}$ is that recommended by \citeauthor{bolatto11a} For comparison, the red dashed line shows the KMT+ model for a Solar metallicity galaxy of the same stellar density at the SMC ($Z'=1$, $\rho_{\rm sd}=0.02$ $\msun$ pc$^{-3}$).
}
\end{figure}

Figure \ref{fig:bolattocomp} shows the comparison between the KMT+ model and \citeauthor{bolatto11a}'s SMC observations. The observational data shown are \citet{bolatto11a}'s 200 pc-resolution data rather than their 12 pc-resolution data, because the latter are severely limited in the range of $\Sigma_{\rm H_2}$ they cover due to signal to noise ratio issues; however, in the limited range where the 12 pc data does exist, it is generally consistent with the 200 pc data (see Figure 5 of \citeauthor{bolatto11a}). As is apparent from the plot, the KMT+ model provides an accurate prediction of the dependence of both $R_{\rm H_2}$ and $\dot{\Sigma}_*$ on the total gas surface density. Notably, the predictions at SMC metallicity are very different from those at Solar metallicity, and the data are a good match to the SMC rather than the Solar metallicity model. Thus the model correctly captures the metallicity-dependence of both the star formation law and the H$_2$ to H~\textsc{i} ratio in the SMC.

\subsection{Nearby Blue Compact Dwarf Galaxies}

The fourth comparison is to a set of nearby blue compact dwarf galaxies compiled by \citet{fumagalli10a}. These galaxies were selected for their low metallicities, and a number of them have high resolution 21 cm maps from which we can extract the peak H~\textsc{i} column density at $\sim 100$ pc or smaller scales. These galaxies provide a useful test of the H~\textsc{i} saturation column predicted by the model. Due to their low metallicities, we expect that the peak H~\textsc{i} column densities of these galaxies should be able to significantly exceed the $\sim 10$ $\msun$ pc$^{-2}$ value found at Solar metallicity, but should stay below the maximum predicted by equation (\ref{eq:himax}).

\begin{figure}
\begin{center}
\includegraphics[width=3.5in]{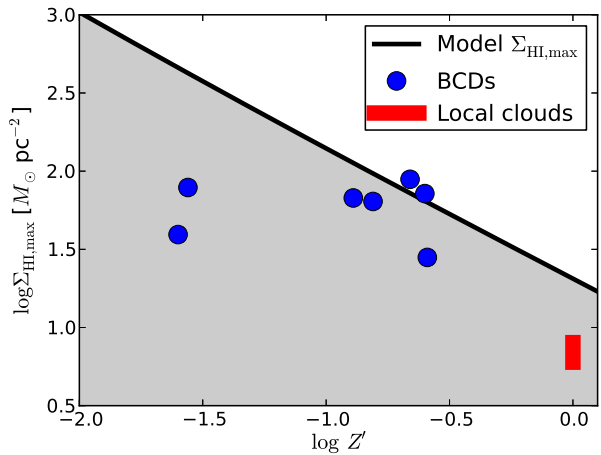}
\end{center}
\caption{
\label{fig:fumagallicomp}
Comparison between the model and the data on blue compact dwarf (BCD) galaxies gathered by \citet{fumagalli10a}. The black line shows $\Sigma_{\rm HI,max}$ (equation \ref{eq:himax}), the metallicity-dependent H~\textsc{i} saturation column, and the gray region below it is the allowed range of maximum H~\textsc{i} column densities. Blue points represent the metallicities and peak H~\textsc{i} columns measured in BCDs by \citeauthor{fumagalli10a}. The red band is the range of H~\textsc{i} saturation columns measured in molecular clouds near the Sun by \citet{lee12a}.
}
\end{figure}

In Figure \ref{fig:fumagallicomp}, I plot the metallicity versus peak H~\textsc{i} column density measured by \citet{fumagalli10a} for local blue compact dwarf galaxies. The behavior of the data is consistent with the model: the peak H~\textsc{i} column densities in the low metallicity galaxies significantly exceed the saturation values measured in the Milky Way, but are below (within the errors) the maximum value predicted by equation (\ref{eq:himax}). For comparison, I also show the H~\textsc{i} saturation column of $6-8$ $\msun$ pc$^{-2}$ measured for Milky Way molecular clouds near by the Sun by \citet{lee12a}. These column densities may be underestimated by a factor of $\sim 2$ due to H~\textsc{i} self-absorption, which would bring them closer to the maximum predicted by the model, but even without this correction the saturation column is a reasonable match to the predicted value, and is far smaller than what is observed in the low-metallicity galaxies.

\subsection{High Redshift Systems}

The final comparison data set consists of damped Lyman $\alpha$ absorbers (DLAs) and the outskirts of Lyman break galaxies (LBGs) at $z\sim 3$. \citet{wolfe06a} combine Hubble Ultra-Deep Field images with Lyman $\alpha$ absorption covering fraction measurements to set upper limits on star formation rates in DLAs, while \citet{rafelski11a} use deep V-band imaging (rest frame FUV) to measure the star formation rate in LBG outskirts, and they show that this star formation, when it can be detected, must be taking place in an H~\textsc{i}-dominated phase of the ISM. By stacking the observed galaxies and statistically comparing to the H~\textsc{i} column density distribution as probed by Lyman $\alpha$ absorption, they are able to determine the connection between the surface densities of star formation and atomic gas.

\begin{figure}
\begin{center}
\includegraphics[width=3.5in]{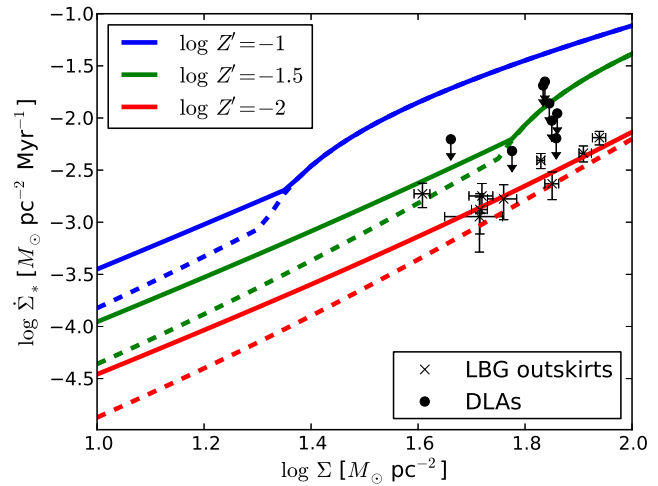}
\end{center}
\caption{
\label{fig:rafelskicomp}
Comparison between the model and observations of damped Lyman $\alpha$ absorbers (DLAs; \citealt{wolfe06a}) and Lyman break galaxy (LBG) outskirts \citep{rafelski11a} at $z\sim 3$ by \citet{rafelski11a}. Black circles and x's show observed values and upper limits, while solid and dashed lines show our models computed with $\log Z'=-1$ (blue), $\log Z'=-1.5$ (green), and $\log Z'=-2$ (red). Solid lines use a stellar density $\rho_{\rm sd} = 0.1$ $\msun$ pc$^{-3}$, while dashed lines use $\rho_{\rm sd} = 0.01$ $\msun$ pc$^{-3}$.
}
\end{figure}

Figure \ref{fig:rafelskicomp} shows a comparison between the measured star formation rates in LBG outskirts, upper limits from DLAs, and the KMT+ model. In generating the model predictions, the choice of metallicity is somewhat unclear, because metallicities are not known on a system-by-system basis. However, \citet{prochaska03a} and \citet{rafelski12a} show that the majority of DLAs at $z\sim 3$ have metallicities in the range $Z' = 0.01 - 0.1$, so we adopt this range, though there are outliers above and below it. Stellar densities are similarly unknown, but have relatively little impact for reasonable values because the gas surface densities are high enough so that stellar gravity only makes a minor contribution to the pressure. The Figure shows that the model agrees reasonably well with the observations for metallicities in the plausible range. Note that the star formation rates are far below what one would expect for an H$_2$-dominated region: total gas depletion times for the LBG outskirts shown are in the range $10-50$ Gyr.

\section{Discussion}
\label{sec:discussion}

\subsection{Prescription for Numerical Simulations and Semi-Analytic Models}

The original KMT model has been adopted as a subgrid model for star formation in a large number of simulations and semi-analytic models that are not able to resolve, or only barely resolve, the phase structure of the ISM \citep[e.g.][]{fu10a, lagos11a, kuhlen12a, forbes12a, jaacks13a, thompson13a, kuhlen13a, forbes13a}. Since the KMT+ model presented here provides a more accurate description of the behavior of gas in the H~\textsc{i}-dominated regime, in this section I describe how to extend a subgrid recipe based on the original KMT model to use KMT+.

For analytic and semi-analytic models, and for simulations that adopt the thin-disk limit, it is natural to phrase a subgrid recipe in terms of the column density, as I do in this paper. The most accurate option in this case is simply to solve the non-linear equations describing the model numerically. However, one may also use the analytic approximation given by equation (\ref{eq:tdepanalyt}), which are nearly as accurate and much faster to evaluate.

For 3D simulations, one most directly has access to volumetric quantities, and thus some effort is required to estimate the column densities that enter the model. The standard approach, introduced by \citet{gnedin09a}, is to use a Sobolev-like approximation to estimate the column density from the local density $\rho$ and its gradient
\begin{equation}
\label{eq:sigmasob}
\Sigma \approx \frac{\rho^2}{|\nabla\rho|}.
\end{equation}
\citeauthor{gnedin09a}~show that this approximation provides a reasonably good estimate of the true column density one would obtain from a ray-tracing procedure. \citet{krumholz11a} show that using the column density estimated in this matter in order to estimate $\tau_c$ in the KMT formalism provides good agreement with the numerical simulations. Thus as a numerical implementation of the KMT+ model, one should use equation (\ref{eq:sigmasob}) to estimate $\Sigma$, and this together with the metallicity to estimate $\tau_c$ from equation (\ref{eq:tauc}).

The other quantity required to compute $f_{\rm H_2}$ using equation (\ref{eq:fH2}) is the normalized radiation field $\chi$. Under the assumption of two-phase equilibrium this depends on the metallicity alone, but here we have relaxed that assumption at low star formation rate. However, computation of $\chi$ is still straightforward, since the true value of $\chi$ is simply the minimum of the values expected for two-phase equilibrium and for hydrostatic balance. The former is given by equation (\ref{eq:chi2p}) and depends on metallicity alone; this part of the computation is the same as for the original KMT model. To obtain the value of $\chi$ if the CNM is at its floor density, one can compute $n_{\rm CNM,hydro}$ from equation (\ref{eq:ncnmhd}). This requires knowledge of the thermal pressure, but in a 3D simulation this is generally known. Computing the normalized radiation field $\chi_{\rm hydro}$ using equation (\ref{eq:chi}) is somewhat trickier, as this requires knowledge of the FUV radiation field. While some simulations include an explicit calculation of $G_0'$ \citep[e.g.][]{gnedin09a}, most do not. The best option will then depend on how the simulation treats stars. In simulations that include star particles for which ages are explicitly tracked, one can estimate $G_0'$ from the local density of young stars simply by scaling from the Milky Way value using equation (\ref{eq:g0}). In simulations that do not track young stars, one can instead compute a self-consistent estimate of $G_0'$ by varying $G_0'$ and thus $\chi_{\rm hydro}$ and $f_{\rm H_2}$ until the value of the star formation rate is consistent with $G_0'$. In any event, once a value of $G_0'$ has been determined by one of these procedures, one can compute $\chi_{\rm hydro}$, the value of $\chi$ one would obtain if the gas were at the minimum density required by hydrostatic balance. One can then compute $f_{\rm H_2}$ from equation (\ref{eq:fH2}) using the smaller of $\chi_{\rm 2p}$ and $\chi_{\rm hydro}$. The result will be a modified estimate of $f_{\rm H_2}$, which can then be fed into an H$_2$-dependent star formation recipe.

\subsection{Implications for Cosmological Star Formation}

In the present-day Universe, the H$_2$-poor mode of star formation we have investigated in this paper exists only in low-metallicity dwarf galaxies and in the outer parts of large spirals. In the early Universe, however, metallicities were much lower, and this mode of star formation was likely to have been more prevalent. A number of authors have investigated the implications of H$_2$-regulated star formation over cosmological times \citep{robertson08b, gnedin09a, gnedin10a, gnedin11a, fu10a, lagos11a, kuhlen12a, christensen12a, krumholz12d, tassis12a, jaacks13a, thompson13a, kuhlen13a} using a variety of strategies, and for those that are based on models such as the original KMT model that have a sharp cutoff in star formation at low metallicity and column density, the results may be changed at least slightly by the updated model, where the cutoff in star formation in the H$_2$-poor regime is more gradual. While this will have to be investigated on a model by model basis, we can make two general observations about the likely implications of our results. 

First, the KMT+ model presented here is unlikely to make a large difference for predictions of global star formation rate of the Universe or similar large-scale quantities. Although KMT+ predicts more star formation in low column density, metal poor galaxies than the original KMT model, the characteristic depletion times associated with this star formation are $\sim 100$ Gyr, longer than a Hubble time. As a result, star formation in this low surface density mode is unlikely to contribute much to the star formation budget of the Universe.

On the other hand, more rapid star formation in metal-poor gas may well affect properties of individual small galaxies, and luminosity and mass functions at the low luminosity end. A common feature of the H$_2$-regulated models that have been calculated thus far is bimodality: some galaxies rapidly self-enrich with metals, and their star formation becomes mostly in the H$_2$ rich regime, while others do not self-regulate and trickle along in the H$_2$ poor regime. To some extent this bimodality is real, and reflects the real bimodality visible in the observations of \citet{bigiel10a} shown in Figure \ref{fig:bigielcomp}: some galaxies really do have depletion times of $\sim 2$ Gyr, while others have depletion times closer to $20-200$ Gyr, depending on the gas column density. However, the previous KMT model overemphasized this bimodality by effectively making the depletion in the H$_2$-poor regime infinity. It seems likely that updating cosmological calculations that used that previously used the KMT model to use the model we present here instead would produce less bimodality.

\subsection{Comparison to the OML Model}
\label{sec:omlcomparison}

The KMT+ model is able to explain a very wide range of observations both nearby and at high redshift. It is interesting to compare these results to the alternative model proposed by OML, which also very successfully matched the original THINGS survey of inner galaxies \citep{bigiel08a}. As discussed above, the OML model is based on the idea that the ISM consists of three neutral components: a warm atomic phase, a cold atomic phase, and a gravitationally-bound phase. Star formation occurs only in the bound phase, with a constant depletion time of 2 Gyr, which is taken from observations; as noted above in local dwarf and disk galaxies, this timescale is very similar to that implied by equations (\ref{eq:sfr}) and (\ref{eq:tff}). The model differs from the one presented here in that there is no explicit treatment of the atomic to molecular transition, and star formation is assumed to follow the gravitationally-bound phase regardless of its chemical state. One computes the mass fraction in the gravitationally-bound phase via a pressure balance argument. The key assumption in this argument, and one that differs significantly from the discussion presented in Section \ref{sec:model}, is that the atomic phases are everywhere in two-phase equilibrium, so that the pressure is proportional to the local radiation field, $G_0'$, and thus to the star formation rate. One then determines the mass fraction in the gravitationally-bound phase that is required to give a star formation rate such that the two-phase pressure is equal to the midplane pressure.

The differing assumptions between the OML and KMT+ models have important implications for their predictions about the behavior of H~\textsc{i}-dominated systems with low surface densities of star formation. First, in the OML model, the star formation rate is effectively set by the weight of the atomic ISM, and thus it is, to good approximation, independent of the metallicity (see OML's equation 22). Second, in the OML model the surface density of $\sim 10$ $\msun$ pc$^{-2}$ where the THINGS data show a turn-down in the star formation rate is not directly due to the atomic to molecular transition, as in the KMT model, but instead is associated with a transition from a region where most of the neutral ISM is diffuse to one where most of it is locked in gravitationally bound clouds. This transition is driven by the balance between stellar gravity and star formation feedback, but because everything behaves smoothly, the result is more a kink at 10 $\msun$ pc$^{-2}$ than a sharp transition in star formation rate per unit mass.

\citet{bolatto11a} show that the OML model provides a poor fit to observations of the SMC, which show a steep drop in star formation rate at an H~\textsc{i} surface density significantly higher than the 10 $\msun$ pc$^{-2}$ predicted by OML. In order to fix this disagreement, they explore a modified version of the model. This modification amounts to adopting an additional assumption that the scaling between FUV radiation field and star formation rate scales inversely with metallicity, so that the newborn stars in the SMC contribute $\sim 5$ times as many FUV photons to the ISRF as do those in the Milky Way. Mathematically, this modification is implemented by altering equation (\ref{eq:g0}) to read
\begin{equation}
G_0' = \frac{1}{Z'}\left(\frac{\dot{\Sigma}_*}{\dot{\Sigma}_{*,0}}\right).
\end{equation}
The physical effect responsible for the extra FUV production is not fully specified, and it is not clear if the modification should apply only to the SMC or to all other galaxies of similar metallicity, but \citeauthor{bolatto11a}~suggest that it might result from reduced dust extinction near stellar birth sites. If this conjecture is correct and applies to low-metallicity galaxies in general, then it should apply to the KMT+ model as well. However, as discussed in Appendix \ref{app:omlh}, the specific mechanism proposed in \citeauthor{bolatto11a}~does not appear to work. I therefore leave equation (\ref{eq:g0}) in its unmodified form for purposes of computing the KMT+ model, but for completeness in this Section I compare to both the original OML model and to the \citet{bolatto11a} modification thereof, which I refer to as OMLZ.\footnote{In \citet{bolatto11a} this model is referred to as OMLh, with the h indicating the proposed extra heating. I refer to it as OMLZ to emphasize the added metallicity-dependence, and because from their discussion it is not clear if the extra heating proposed in the OMLh model is purely a metallicity effect or is produced by some other property of the SMC as well. For OMLZ, I explicitly assume that the enhancement is purely a metallicity effect.}

\begin{figure*}
\begin{center}
\includegraphics[width=6.0in]{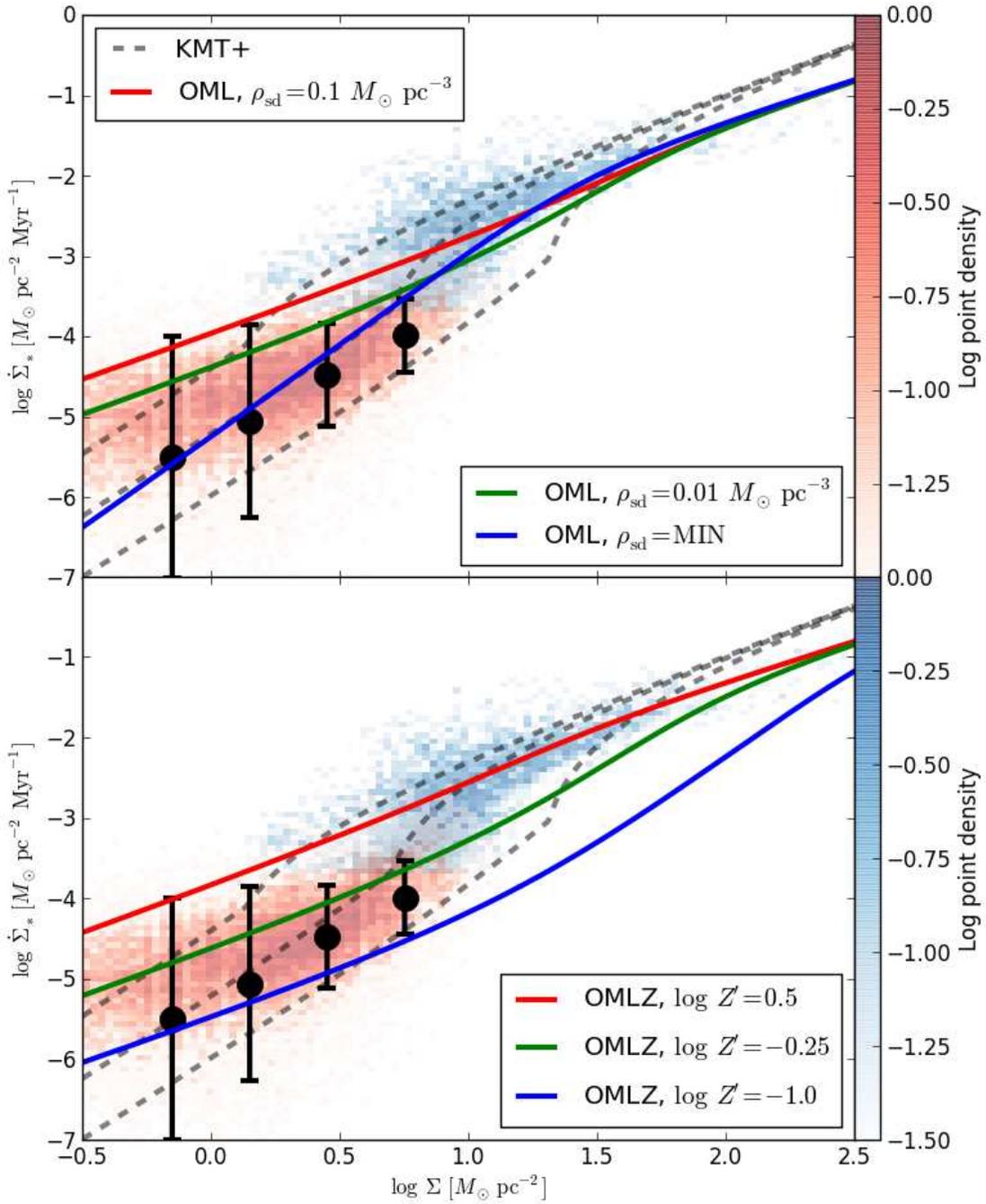}
\end{center}
\caption{
\label{fig:bigiel_oml}
Same as Figure \ref{fig:bigielcomp}, but now showing the OML (top panel) and OMLZ (bottom panel) models over-plotted on the observations of \citet{bigiel08a, bigiel10a}. In the upper panel, the red, green, and blue solid lines show the OML model computed with $\rho_{\rm sd} = 0.1$ $\msun$ pc$^{-3}$, $0.01$ $\msun$ pc$^{-3}$, and the minimum possible value of $\rho_{\rm sd}$ given by equation (\ref{eq:rhosdmin}). All evaluations use $Z'=1$, but this choice has minimal effects for the OML model. In the lower panel, the red green, and blue solid lines show the OMLZ model computed for metallicities of $\log Z'=-1.0$, $-0.25$, and $0.5$, and $\rho_{\rm sd} = 0.01$ $\msun$ pc$^{-3}$. In both panels, the gray dashed lines show the KMT+ model computed for the same metallicity and stellar density as the OMLZ models. Note that the gray dashed lines here are identical to the dashed lines in Figure \ref{fig:bigielcomp}.
}
\end{figure*}

\begin{figure}
\begin{center}
\includegraphics[width=3.5in]{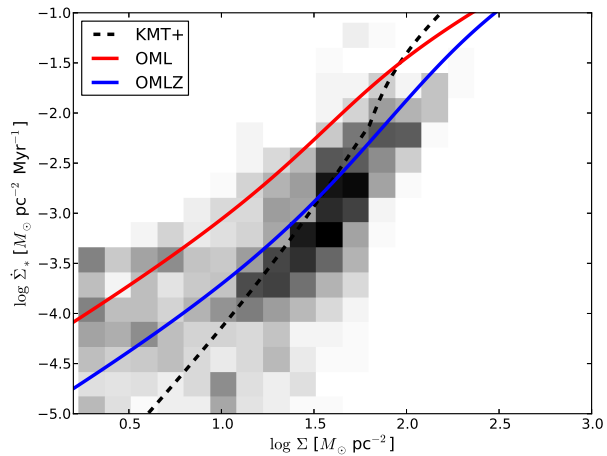}
\end{center}
\caption{
\label{fig:bolatto_oml}
Same as Figure \ref{fig:bolattocomp}, but now showing the OML (red) and OMLZ (blue) models over-plotted on the observations of \citet{bolatto11a}. The black dashed line is the KMT+ model.
}
\end{figure} 

\begin{figure}
\begin{center}
\includegraphics[width=3.5in]{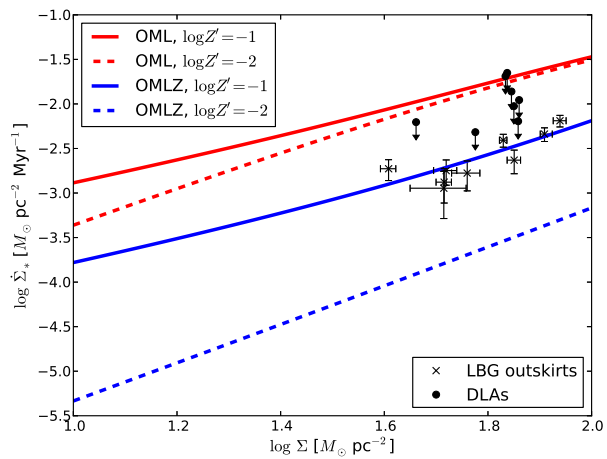}
\end{center}
\caption{
\label{fig:rafelski_oml}
Same as Figure \ref{fig:rafelskicomp}, but now showing the OML (red) and OMLZ (blue) models over-plotted on the observations of \citet{rafelski11a}. Solid lines show $\log Z'=-1$ and $\rho_{\rm sd} = 0.1$ $\msun$ pc$^{-3}$, while dashed lines show $\log Z'=-2$ and $\rho_{\rm sd}$ set equal to the minimum possible value (equation \ref{eq:rhosdmin}).
}
\end{figure}

To see how the OML and OMLZ models compare to the KMT+ model, Figures \ref{fig:bigiel_oml} --  \ref{fig:rafelski_oml} show them compared to a subset of the observational data sets discussed in the previous section. The Figures do not include comparisons to data sets that separate the ISM into H~\textsc{i} and H$_2$, since the OML model does not explicitly consider the partition of the ISM between these two phases. For the purposes of all these plots, I evaluate the OML model by numerically solving equations (5) -- (9) of \citet{bolatto11a}, and the OMLZ model via the same procedure but with their equation (8) in place of their equation (11). 

Examining the figures, it is clear that the original OML is in some tension with the observed star formation rates of galaxies in the H~\textsc{i}-dominated regime. While \citet{bolatto11a} had already demonstrated this for the SMC (as illustrated in Figure \ref{fig:bolatto_oml}), the problem also appears elsewhere. Comparing to the \citet{bigiel08a, bigiel10a} observations (Figure \ref{fig:bigiel_oml}) shows that, even if one sets $\rho_{\rm sd}$ equal to the value corresponding to a disk of gas and dark matter only with no stars (which is probably close to the appropriate comparison in outer disks), the \textit{minimum} star formation rate predicted by the OML model is at or above the \textit{median} value in observed galaxies, and exceeds the bottom of the $1\sigma$ range (the lower ends of the black error bars) by more than an order of magnitude. One could marginally improve the agreement by assuming a lower $Q_g$ or a higher $\sigma_g$ than the fiducial values used in equation (\ref{eq:rhosdmin}), but to bring the minimum star formation rate predicted by the OML model down enough to match the data would require fairly drastic choices. At low gas surface densities, the OML model predicts that the star formation rate scales with stellar density as $\dot{\Sigma}_*\propto \rho_{\rm sd}^{1/2} \propto Q_g/\sigma_g$, so the required reduction in $\dot{\Sigma}_*$ could be achieved only by lowering $Q_g$ from 2 to $<0.2$, raising $\sigma_g$ from 8 km s$^{-1}$ to $>80$ km s$^{-1}$, or some combination of the two. Figure \ref{fig:rafelski_oml} shows that a similar tension exists between the OML model and the SFRs observed in high redshift H~\textsc{i}-dominated systems, even if one again sets $\rho_{\rm sd}$ to its minimum value. In this case the value of $\rho_{\rm sd}$ makes little difference, because the gas surface densities are so large that gas self-gravity dominates over stellar gravity.

The OMLZ model provides a significantly better fit to the observations, and generally makes predictions that are similar to those of the KMT+ model. The extra metallicity-dependence allows it to reproduce the observed long depletion times found in low-metallicity, H~\textsc{i}-dominated regions. However, I caution again (see Appendix \ref{app:omlh}) that the physical motivation for this extra metallicity dependence in the OML framework is not clear, and that the property responsible for lowering the star formation rate could be some property other than metallicity that is nevertheless common to the SMC, outer spiral disks, and high-redshift galaxies. If metallicity really is the important variable, however, as assumed in model OMLZ, then Figure \ref{fig:bigiel_oml} suggests a powerful means for discriminating between models of this form and the KMT+ model. The OMLZ and KMT+ models predict similarly low star formation rates for low-metallicity, low surface density systems, but for KMT+ metallicity depresses the star formation rate only at low surface density (black dashed lines in Figure \ref{fig:bigiel_oml}), while in OMLZ or a similar model metallicity reduces the star formation rate at all surface densities (blue line in Figure \ref{fig:bigiel_oml}).

The physical origin of this disagreement is easy to understand. In the KMT+ model, star formation cares about metallicity because dust grains mediate the H~\textsc{i} to H$_2$ transition. This is primarily a shielding effect: one requires a certain metallicity-dependent column of gas before the ISM transitions from H$_2$-poor to H$_2$-rich, but at surface densities above this shielding column, star formation does not behave any differently than it would in a higher metallicity galaxy with the same gas column density. This is why the KMT+ model curves at different metallicities all converge at high $\Sigma$ (see Figures \ref{fig:bigielcomp} and \ref{fig:bigiel_oml}). In the OMLZ model, on the other hand, the convergence of models of differing metallicity at high surface densities is much slower or absent. For the specific example of OMLZ, this is because a lower metallicity raises the amount of heating per unit star formation, and thus depresses the star formation rate at all surface densities. More generally, we can distinguish between models like KMT+ where metallicity matters primarily through dust shielding effects, and models like OMLZ where metallicity affects star formation in some other way. Shielding-based models predict the metallicity ceases to matter at high surface densities, while non-shielding ones predict that metallicity affects star formation at all surface densities, not just those below some threshold.

Actually performing a test to distinguish these models requires knowledge of the stellar density as well, since that also affects the star formation rate in the OMLZ model. Nonetheless, we can state the required test very simply: consider two galaxies, or regions within them, that both have $\Sigma_g \approx 100$ $\msun$ pc$^{-2}$, and that have equal stellar densities, but one of which has $Z' = 1$ and the other of which $Z'=0.1$. Examining Figure \ref{fig:bigiel_oml}, we see that the KMT+ model predicts that these two galaxies will have roughly the same star formation rate per unit area, while the OMLZ model predicts that their star formation rates per unit area will differ by an order of magnitude. Unfortunately the SMC data does not quite reach the required surface densities and metallicities to constitute a strong test, since the data run out just about where the divergence between KMT+ and OMLZ begins.

\section{Summary}
\label{sec:summary}

I present a new analytic model for the atomic to molecular transition and the star formation law in the outer regions of spiral galaxies and in low-metallicity dwarf galaxies. Observations of these systems show that the ISM in these regions is composed primarily of non-star-forming H~\textsc{i}, and that the star formation rates per unit total gas mass is $\sim 1-2$ orders of magnitude smaller than in molecule-rich inner, metal-rich parts of spiral galaxies. This model extends the formalism developed by \citet[KMT]{krumholz08c, krumholz09a, krumholz09b} to provide more accurate results in the H~\textsc{i}-dominated regime.

The central idea behind this ``KMT+" model is to consider what processes set the density of the cold neutral gas that mediates the transition from inert, warm H~\textsc{i} to very cold, star-forming H$_2$. In the inner parts of galaxies, the atomic ISM is in two-phase equilibrium, and in this state the cold gas density and the local FUV radiation field are approximately proportional to one another. However, this state cannot continue to hold in once the star formation rate is too small, because at that point the density at which the cold atomic gas could be in two-phase equilibrium is less than the minimum density that is required for this gas to be in hydrostatic balance against the weight of the galactic disk. When this condition holds, it is a reasonable hypothesis that the density of the cold atomic gas will be roughly equal to the minimum imposed by hydrostatic balance. This assumption makes it possible to compute the H$_2$ fraction and the star formation rate as a function of gas surface density, metallicity, and the density of the stellar disk.

I show that the star formation rate and H$_2$ fraction computed following this procedure naturally explain a number of previously puzzling observations. The model naturally explains why the star formation rate per unit area in galactic disks suddenly drops by $\sim 1 - 2$ orders of magnitude at a critical, metallicity-dependent gas column density, and it correctly and quantitatively predicts both the critical column density and the star formation rate for gas that is below this critical column density. The model is able to reproduce resolved observations of nearby spiral and dwarf galaxies, as well as local low metallicity systems such as the the Small Magellanic Cloud and blue compact dwarf galaxies. It is also able to reproduce the statistically-inferred relationship between star formation and H~\textsc{i} surface density at $z\sim 3$. Finally, I provide a recipe for implementing this model as a subgrid recipe for star formation in simulations and semi-analytic models of galaxy formation.

\section*{Acknowledgements}

I thank F.~Bigiel, A.~Bolatto, M.~Rafelski, A.~Schruba, and F.~Walter for providing their observational data, and for advice in interpreting it. I thank C.~McKee and E.~Ostriker for helpful discussions and comments on the manuscript. This work was supported by the Alfred P.~Sloan Foundation, the NSF through CAREER grant AST-0955300, and NASA through ATP grant NNX13AB84G. I thank the Aspen Center for Physics, which is supported by NSF Grant 1066293, for hospitality during the writing of this paper.

\bibliographystyle{mn2e}
\bibliography{refs}

\begin{appendix}

\section{The Thermal Velocity Dispersion}
\label{app:fw}

As discussed in the main text, the value of $\tilde{f}_w$, the ratio of the mass-weighted mean square thermal velocity dispersion to the square of the warm gas sound speed, is quite uncertain. If there is significant WNM present, since its sound speed is much greater than that of the CNM, $\tilde{f}_w$ is nearly identical to the mass fraction in the warm phase. Based on this consideration, and the fact that WNM is observed to be present over a wide range of galactic radii (though not necessarily at the midplane, which is what matters for this purose), OML adopt $\tilde{f}_w = 0.5$ as a fiducial value. Simulations appear to be roughly consistent with this value, at least at surface densities $\sim 3$ $\msun$ pc$^{-2}$ or more \citep{kim11a}.
However, in the very low surface density, outer disk regions that are our concern in this paper, the ISRF is very weak, and the pressure is such that there is not expected to be a stable warm phase at the midplane, though one should exist at larger altitudes. If the WNM fraction at the midplane were truly zero then we would have $\tilde{f}_w \approx 0.01$, since the squared velocity dispersion on the CNM is roughly 1\% that of the WNM. However, such a low value seems quite improbable. Observations show that, even in regions where both stable warm and cold phases are stable, a significant amount of mass is nonetheless in the unstable intermediate regime \citep{jenkins01a, jenkins11a, heiles03a}. The same is likely to be true in the outer disk regions we are interested and modeling, and if even a small amount of such non-equilibrium gas is present, this would be sufficient to raise $\tilde{f}_w$ well above $0.01$. For example, \citet{heiles03a} estimates that $\sim 30\%$ of the H~\textsc{i} in the Solar neighborhood is at unstable temperatures of $500 - 5000$ K; adopting a temperature of 1600 K for this gas, the geometric mean of the two limits, this would give $\tilde{f}_w = 0.05$ even if there were no WNM present. If any WNM penetrates to the midplane, even though it is not expected to be stable there, $\tilde{f}_w$ would be even higher. Unfortunately the simulations that have been done thus far \citep[e.g.,][]{kim11a}, while they show some decline in $\tilde{f}_w$ with gas surface density, have not probed into the far outer disk regime where the warm phase ceases to be stable at the midplane.

\begin{figure*}
\begin{center}
\includegraphics[width=6.0in]{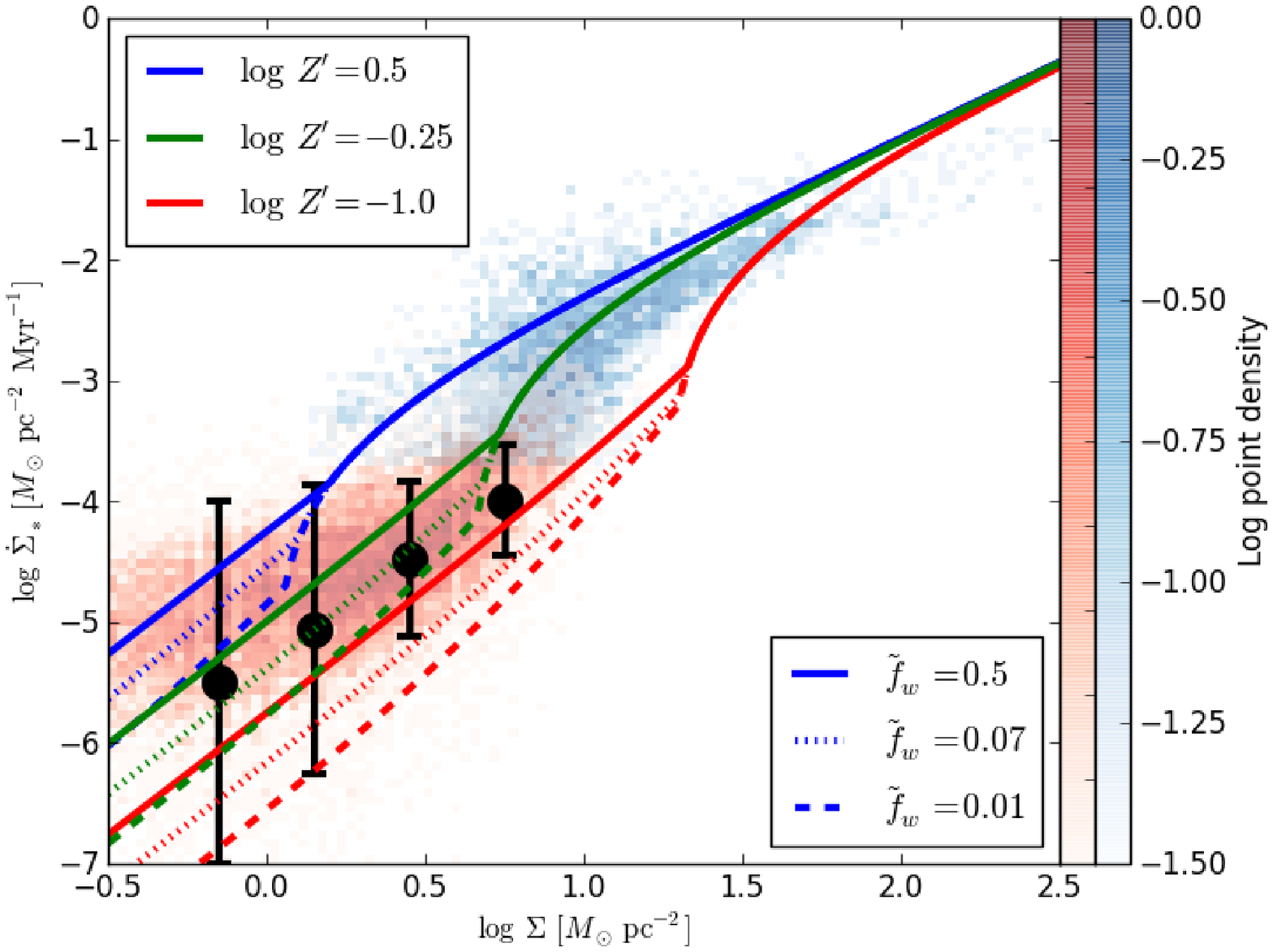}
\end{center}
\caption{
\label{fig:bigiel_fw}
Same as Figure \ref{fig:bigielcomp}, but showing models for different values of $\tilde{f}_w$. Solid lines show models with our fiducial value, $\tilde{f}_w$, while dotted and dashed lines show $\tilde{f}_w = 0.07$ and $\tilde{f}_w = 0.01$, respectively. Red, green, and blue lines correspond to metallicities of $\log Z' = 0.5, -0.25$, and $-1.0$. All the curves shown use $\rho_{\rm sd} = 0.03$ $\msun$ pc$^{-3}$ and $f_c = 5$. The background raster plot shows the observations of \citet{bigiel10a}; see the caption to Figure \ref{fig:bigielcomp} for details.
}
\end{figure*}

Given the uncertainties, and for the sake of consistent comparison with OML, I adopt $\tilde{f}_w \approx 0.5$ as a fiducial value. To explore the implications of a lower value, in Figure \ref{fig:bigiel_fw} I compare models with different values of $\tilde{f}_w$, overplotted on the observations of \citet{bigiel10a}. We see that, unsurprisingly given the form of equation (\ref{eq:pth}), changes in $\tilde{f}_w$ are essentially degenerate with changes in $\rho_{\rm sd}$, and that essentially all the plotted values of $\tilde{f}_w$ produce model curves within the (very large) scatter of the observations. Physically, different values of $\tilde{f}_w$ correspond to different assumptions about the scale height of the gas layer, and thus to differing assumptions about the amount of stellar matter within the gas layer and contributing to its gravity. A value of $\tilde{f}_w = 0.01$ would correspond to a gas layer in which the CNM and WNM are perfectly stably stratified, with CNM at the midplane and WNM above at heights where it can exist stably, while $\tilde{f}_w$ corresponds to assuming that, despite the fact that WNM is not stable at the midplane, enough WNM or unstable gas reaches the midplane to stir up the CNM and raise its scale height, making stellar gravity more important for it.

\section{Scaling Between FUV Radiation Field and Star Formation Rate}
\label{app:omlh}

\citet{bolatto11a} propose that newborn stars in the Small Magellanic Cloud contribute significantly more FUV photons to the ISRF than do stars of similar properties in the Milky Way. The enhancement required for the OML model to match the observations of the SMC is a factor of $\sim 5$, although if the SMC is highly inclined then a smaller enhancement may be sufficient. If enhanced FUV production is a general features of low metallicity systems, it will be important for the KMT+ model as well. In this Appendix I argue, however, such an enhancement is unlikely. Changes in the intrinsic colors of young stars with metallicity enhance the FUV photon production rate by significantly less than this amount, and both the photoelectric heating efficiency per unit metal mass and the escape of FUV radiation from a galaxy and into intergalactic space goes in the wrong direction, in the sense that one would expect less efficient photoelectric heating and more photon escape in galaxies like the SMC with lower metal and dust content. The main reason that one might expect to find more heating of the diffuse ISM in low metallicity galaxies is because a higher proportion of the FUV photons escape from the clouds in which they are born and propagate into the diffuse ISM to be absorbed there. This mechanism, dubbed the proximity effect, is the one that \citet{bolatto11a} suggest might be responsible for the enhanced heating required by the model in the SMC.

However, the only way it could be the case that five times as many FUV photons escape from their parent clouds in the SMC as do in the Solar neighborhood would be if the escape fraction of FUV photons from their parent clouds in the Solar neighborhood were $\leq 20\%$. Indeed, \citet{bolatto11a} estimate an upper limit on the proximity effect by assuming an escape fraction of 0 for the Milky Way (equation 13 of \citeauthor{bolatto11a}~and surrounding discussion). However, even a 20\% escape fraction for Milky Way-like galaxies faces two severe problems. First, observationally-estimated FUV extinctions are generally too low to be consistent with this requirement. An escape fraction of 20\% corresponds to local clouds typically having $1.75$ magnitudes of absorption in the UV; the corresponding FUV attenuation we should observe from a Milky Way-like galaxy would then be $2-3$ magnitudes, both because the extinction coefficient is $\sim 50\%$ larger than the absorption coefficient at FUV wavelengths, and because there will also inevitably be some extinction that is not local to the parent molecular cloud. In contrast, \citet{boissier07a} find that Solar metallicity galaxies with surface densities of $\sim 10$ $\msun$ pc$^{-2}$ generally show $\sim 1$ mag of attenuation in the FUV.

Second, there is a timescale problem: FUV photons are produced over a timescale of $10-100$ Myr, while stars typically remain embedded in their parent molecular clouds for $< 10$ Myr. To be precise, \citet{kawamura09a} find that, after roughly 7 Myr of age, there is no longer a statistically-discernible correlation between the positions of star clusters and molecular clouds in the Large Magellanic Cloud. This sets a firm upper limit on the length of time for which star clusters could potentially have their FUV output absorbed by their birth clouds. However, a Starburst99 \citep{leitherer99a, vazquez05a} calculation for an instantaneous burst of star formation indicates that roughly $1/3$ of the luminosity at 1000 \AA~emerges \textit{after} 7 Myr, implying that the proximity effect could remove at most $2/3$ of the FUV photons produced in the LMC. This is certainly a large overestimate of the true proximity effect, because this corresponds to the assumption that the FUV escape fraction is identically zero at ages $< 7$ Myr, which is manifestly false.

Thus a realistic assessment of the proximity effect in the Solar neighborhood is that it reduces FUV heating of the diffuse ISM by at most a factor of $\sim 2$. This in turn implies that a factor of $\sim 2$ is the maximum possible amount by which FUV photon production in the SMC could be enhanced relative to the Milky Way.

\end{appendix}

\end{document}